\documentclass{statsoc}

\usepackage[a4paper]{geometry}
\usepackage{graphicx}
\usepackage[textwidth=8em,textsize=small]{todonotes}
\usepackage{natbib}
\usepackage{soul}
\usepackage[caption=false]{subfig}
\usepackage{amsmath}
\usepackage{amssymb}
\usepackage{mathtools}
\usepackage{bbm}
\usepackage{bm}
\usepackage{algorithm2e}
\usepackage{tabularx}
\usepackage{booktabs}
\usepackage{makecell}
\usepackage{multicol}
\usepackage{url}
\usepackage{tikz}

\newcommand\blfootnote[1]{%
  \begingroup
  \renewcommand\thefootnote{}\footnote{#1}%
  \addtocounter{footnote}{-1}%
  \endgroup
}

\definecolor{RawDenimBlue}{RGB}{59, 91, 146}
\colorlet{DenimBlue}{RawDenimBlue}
\DeclareRobustCommand\DenimBlueFull{\tikz[baseline=-0.6ex, color=DenimBlue]\draw[ultra thick] (0,0)--(0.5,0);}
\DeclareRobustCommand\DenimBlueDashed{\tikz[baseline=-0.6ex, color=DenimBlue]\draw[ultra thick,dashed] (0,0)--(0.5,0);}

\definecolor{RawPaleRed}{RGB}{217, 84, 77}
\colorlet{PaleRed}{RawPaleRed}
\DeclareRobustCommand\PaleRedFull{\tikz[baseline=-0.6ex, color=PaleRed]\draw[ultra thick] (0,0)--(0.5,0);}
\DeclareRobustCommand\PaleRedDashed{\tikz[baseline=-0.6ex, color=PaleRed]\draw[ultra thick, dashed] (0,0)--(0.5,0);}
\DeclareRobustCommand\PaleRedDotted{\tikz[baseline=-0.6ex, color=PaleRed]\draw[ultra thick, dotted] (0,0)--(0.5,0);}

\definecolor{RawBoringGreen}{RGB}{99, 179, 101}
\colorlet{BoringGreen}{RawBoringGreen}
\DeclareRobustCommand\BoringGreenFull{\tikz[baseline=-0.6ex, color=BoringGreen]\draw[ultra thick] (0,0)--(0.5,0);}
\DeclareRobustCommand\BoringGreenDotted{\tikz[baseline=-0.6ex, color=BoringGreen]\draw[ultra thick,dotted] (0,0)--(0.5,0);}
\DeclareRobustCommand\BoringGreenDashed{\tikz[baseline=-0.6ex, color=BoringGreen]\draw[ultra thick,dashed] (0,0)--(0.5,0);}

\definecolor{RawOrange}{RGB}{249, 115, 6}
\colorlet{ModifiedOrange}{RawOrange}
\DeclareRobustCommand\OrangeFull{\tikz[baseline=-0.6ex, color=ModifiedOrange]\draw[ultra thick] (0,0)--(0.5,0);}

\definecolor{RawDustyPurple}{RGB}{130, 95, 135}
\colorlet{DustyPurple}{RawDustyPurple}
\DeclareRobustCommand\DustyPurpleFull{\tikz[baseline=-0.6ex, color=DustyPurple]\draw[ultra thick] (0,0)--(0.5,0);}

\definecolor{RawBrown}{RGB}{101, 55, 0}
\colorlet{ModifiedBrown}{RawBrown}
\DeclareRobustCommand\BrownFull{\tikz[baseline=-0.6ex, color=ModifiedBrown]\draw[ultra thick] (0,0)--(0.5,0);}

\definecolor{RawGrey}{RGB}{146, 149, 145}
\colorlet{ModifiedGrey}{RawGrey}
\DeclareRobustCommand\GreyDashed{\tikz[baseline=-0.6ex, color=ModifiedGrey]\draw[ultra thick,dashed] (0,0)--(0.5,0);}

\DeclareRobustCommand\BlackFull{\tikz[baseline=-0.6ex]\draw[ultra thick] (0,0)--(0.5,0);}
\DeclareRobustCommand\BlackDashed{\tikz[baseline=-0.6ex]\draw[ultra thick,dashed] (0,0)--(0.5,0);}

\title{A non-parametric Hawkes process model of primary and secondary accidents on a UK smart motorway}

\author{Kieran Kalair}
\address{Centre for Complexity Science, 
University of Warwick, 
Coventry,
United Kingdom
}

\author{Colm Connaughton}
\address{London Mathematical Laboratory, 
London, 
United Kingdom

Mathematics Institute,
University of Warwick, 
Coventry, 
United Kingdom
}

\author[K. Kalair, C. Connaughton, P. Alaimo Di Loro]{Pierfrancesco Alaimo Di Loro}
\address{Department of Statistics, University of Rome "La Sapienza", Rome, Italy}

\begin{document}
\begin{abstract}
A self-exciting spatio-temporal point process is fitted to incident data from the UK National Traffic Information Service to model the rates of primary and secondary accidents on the M25 motorway in a 12-month period during 2017-18. 
This process uses a background component to represent primary accidents, and a self-exciting component to represent secondary accidents. 
The background consists of periodic daily and weekly components, a spatial component and a long-term trend. 
The self-exciting components are decaying, unidirectional functions of space and time. 
These components are determined via kernel smoothing and likelihood estimation. 
Temporally, the background is stable across seasons with a daily double peak structure reflecting commuting patterns. 
Spatially, there are two peaks in intensity, one of which becomes more pronounced during the study period. 
Self-excitation accounts for 6-7\% of the data with associated time and length scales around 100 minutes and 1 kilometre respectively.  
In-sample and out-of-sample validation are performed to assess the model fit.  
When we restrict the data to incidents that resulted in large speed drops on the network, the results remain coherent.

\end{abstract}
\blfootnote{ \emph{Address for correspondence:} Kieran Kalair, Centre for Complexity Science, Zeeman Building, University of Warwick, Coventry, CV4 7AL, UK. Email: k.kalair@warwick.ac.uk  }

\section{Introduction}\label{sec:Introduction}

The United Kingdom has one of the lowest per-capita death rates from traffic accidents in the world, estimated by the \cite{world2018global} at 3.1 per 100,000 of population in 2016. 
Nevertheless 1782 deaths and 25,484 serious injuries resulted from accidents on UK roads in 2018 \cite{reported_road_casualities}.
Aside from the direct human cost of serious accidents, indirect economic costs result even from relatively minor incidents.
This is because crashes, collisions and breakdowns can cause severe congestion leading to significant drops in the efficiency of the road transport network.
For these reasons, there is an imperative to further reduce the accident rate on UK roads.
However traffic accidents are rare in absolute terms and are not distributed uniformly across the network. 
Further rate reductions are therefore likely to require targeted interventions. 

Targeted interventions might try to improve safety at specific locations where the accident risk is known to be high compared to the baseline or might try to mitigate against particular mechanisms that are known to account for a significant proportion of accidents. 
Infrastructure modifications to improve the safety of accident-prone junctions is an example of the first type of intervention. 
The deployment of the Motorway Incident Detection and Automatic Signalling (MIDAS) system to reduce the number of secondary accidents on motorways is an example of the second.
Secondary accidents occur when a driver fails to react appropriately to the disruption caused by an existing accident leading to a subsequent accident upstream of the first one. 
MIDAS uses a network of induction sensors, known as loops, embedded in the road surface to automatically detect queues and then warn upstream drivers of the danger ahead via roadside signage, and is one component of the UK `smart motorways' infrastructure, see  \cite{how_to_drive_on_a_smart_motorway} for further details.
Most of the data, including data on accidents and congestion, is publicly available via the National Traffic Information Service (NTIS). 

In this paper, we use data from NTIS to model the distribution of motorway incidents as a spatio-temporal process comprised of a background component and a self-excitation component, see section \ref{sec:Methodology} for details. 
We focus on the M25 London Orbital, one of the busiest motorways in the UK.
The objectives of the study are two-fold. 
The first is to quantify how accident risk on the M25 varies in space and time relative to the baseline. 
The second is to use the self-excitation component of the process to quantify the likely contribution of secondary incidents to the observed totals.
We would like this model to be helpful in addressing the question of how best to target interventions when the baseline accident rate is low in absolute terms.
We therefore perform extensive in-sample and out-of-sample validation to verify the models performance on seen and unseen data.

\section{Literature Review}\label{sec:LiteratureReview}


There is much work focusing on spatial and temporal analysis of traffic incidents, with a discussion of the evidence for spatial auto-correlation in incident data given in \cite{analysis_of_road_crash_frequency_with_spatial_models}.
Here, the authors discussed previously used descriptive analysis methods, including K-function analysis and comparison to complete spatially random patterns, and detail how they showed evidence of spatial correlation among event locations.
They furthered this by incorporating spatial components into an auto-regressive model, finding it improved upon models disregarding the spatial correlation in the data.
Additional analysis is provided in \cite{network_contrained_spatio_temporal_clustering_analysis_of_traffic_collisions_in_jianghan_district_of_wuhan_china}, where the authors analysed the evolution of event hot-spots through time on an urban network in China.
They extended and applied kernel density estimation on networks, Moran's I and Local Indicators of Mobility Associations (LIMA) to offer exploratory analysis of the dataset from multiple perspectives.
Their work was focused on analysis of the data, rather than formulating a model incorporating the observed features, as we do here.

Further spatial-temporal analysis was completed in \cite{identification_of_traffic_accident_clusters_using_kulldorffs_space_time_scan_statistics}, where the authors used Kulldorff's space-time scan statistics to determine statistically significant clusters of traffic accidents across the entire UK in 2016.
They found two significant clusters, both in the north of the country, but conceded that they do not explicitly account for the network structure in their analysis.
We also consider data from the UK, but focus on a single motorway rather than the entire country.
Our approach is fundamentally different as we aim to model the dynamics as a point-process, not just discover locations of statistically significant clusters.
Readers interested in the discussed spatial analysis methodologies can find details of K-function analysis in \cite{ripleys_k_function}, Kulldorff's space-time scan statistic in \cite{prospective_time_periodic_geographical_disease_surveillance_using_a_scan_statistic} and LIMA in \cite{space_time_patterns_of_rank_concordance_local_indicators_of_mobility_association_with_application_to_spatial_income_inequality_dynamics}.

A point of particular interest to our work arises from the modelling of traffic incidents in \cite{network_based_likelihood_modeling_of_event_occurances_in_space_and_time_a_case_study_of_traffic_accidents_in_dallas_texas}.
Here, logistic regression and random forest models were used to predict the likelihood of event occurrences.
In-particular, the models incorporated a significant  `cascading effect' variable, in which the presence of an event showed significant influence on the likelihood of another nearby in space and time.
Although a different type of model and formulation to the one we use, there is clearly a sense that cascading effects are a real component in some traffic data that one may want to incorporate into a model.
It is unclear what time and length scales are associated with this effect however, and if smart motorway features may remove this effect from the data.

Based on this analysis, it is natural to consider applying spatio-temporal point-process models to traffic events. 
Recent work considering these models on linear networks is given in \cite{first_and_second_order_characteristics_of_spatio_temporal_point_processes_on_linear_networks}, with an application to traffic event modelling.
Considering road-networks in Huston, Medellin and Eastbourne, the authors investigated what features the network-constrained data showed.
They found statically significant evidence that events on the network did not follow a uniform spatial-temporal Poisson process and that tests indicated favouring clustering of data in space and time.
Whilst their focus was on urban road networks, ours is on a smart motorway.
As we consider one continuous ring-road, not an urban network, we can also use a standard distance measure, rather than a graph distance based one as in the cited work.
Our choice of dataset is not to simplify the fitting procedure onto continuous space rather than a network, but instead it offers some insight into a much discussed topic of smart motorway safety.


From our review of the literature it is clear that one could ask if point-process models incorporating a self excitation component capture this previously discussed cascading effect.
Such models have been applied to a wide range of real-world problems, with a recent review of these models given in \cite{a_review_of_self_exciting_spatio_temporal_point_processes_and_their_applications}. 
One application discussed in this review is earthquake modelling, as in \cite{analyzing_earthquake_clustering_features_by_using_stochastic_reconstruction} and \cite{next_day_earthquake_forecasts_for_the_japan_region_generated_by_the_etas_model}.
This is a natural application for such models, as there is strong evidence that initial large earthquakes lead to aftershocks, and hence there is a clearly interpretable `self-excitation' component to the application.
Another application discussed is crime forecasting, as in \cite{self_exciting_point_process_modelling_of_crime} and \cite{marked_point_process_hotspot_maps_for_homicide_and_gun_crime_prediction_in_chicago}, where self-excitation can be seen in physical terms as retaliation crimes, among other things.
Alternative applications discussed include epidemic forecasting in \cite{a_recursive_point_process_model_for_infectious_diseases}, and modelling events on social-networks as in \cite{modelling_email_networks_and_inferring_leadership_using_self_exciting_point_processes}. 
Very recently, similar models have been applied to modelling the spread of COVID-19 \cite{a_hawkes_process_to_make_aware_people_of_the_severity_of_covid_19_outbreak_application_to_cases_in_france}, although this is still in its infancy.

Clearly, there is significant work on applying self-exciting point-process models to problems in crime, earthquakes and epidemics, but there is little on applying them to traffic events.
One paper that does look at this is \cite{traffic_accident_modelling_via_self_exciting_point_processes}, where a self-exciting point-process model is proposed that could theoretically be fit to real data.
However, only simulated data is used, generated from the proposed model, to then show how the fitting and evaluation would work.
Additionally, a somewhat similar idea was considered in \cite{traffic_flow_modelling_with_point_processes}, however importantly here the authors considered traffic flow data to be `events' and tried to use self-excitation to model the idea that often traffic flow occurs in clusters.
They then applied the methods to model traffic flow in Sydney, however there was no clear conclusion as to if the model was statistically defensible and captured all features of the data, or if alternative traffic forecasting methods were preferable.
There is still an enormous amount of work to be done applying this methodology to real-data, and understanding what components of it are important, and the amount of self excitation present in traffic data, along with appropriate time and length scales it occurs on.
Throughout our modelling, we are able to offer some discussions of these questions. 

In the discussed works using self-exciting point processes, there is some sense of a `background' component, that models the typical behaviour, and a `triggering' component that allows for self-excitation.
There is much discussion as to what functional form the components should take, in-particular for the triggering components.
Typically, one supposes some reasonable parametric forms, then determines which is most appropriate through inspection of the log-likelihood value or information criteria.
Standard choices include some form of exponential, Gaussian or power-law decay of triggering in time and space.
However, recent work in \cite{a_semiparametric_spatiotemporal_Hawkes_type_point_process_model_with_periodic_background_for_crime_data} showed how one could determine both the background and triggering components in a non-parametric way, through kernel-smoothing of data.
In-particular, the authors proposed to model crime data using a background comprised of periodic daily, periodic weekly, long-term trend and spatial components, as-well as triggering in space and time.
However, every component is determined without assuming any functional form, instead the authors show how when basing their methods on work in \cite{second_order_residual_analysis_of_spatiotemporal_point_processes_and_applications_in_model_evaluation}, one can determine which events in the data appear to be background and which appear to be triggered, and then smooth data based on this to reconstruct the desired components.
It is on this we base most of our work in this paper, and in section \ref{sec:Methodology} we give an overview of the methodology and how we adapt it for our use.

\section{Data Collection and Pre-Processing}\label{sec:Data}

\subsection{Collection \& Selection}
\begin{figure}[!ht]
	\centering
	\includegraphics[width=0.48\textwidth]{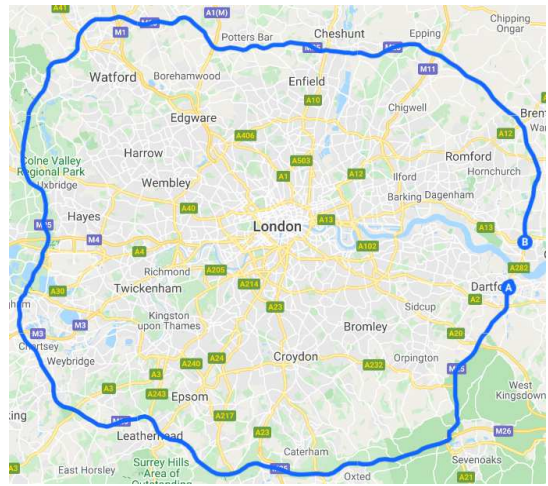}
	\caption{The M25 London Orbital, roughly 180 kilometres in length. The Dartford crossing, located in the east, is a short segment of road that we do not have data for.}\label{fig:M25Plot}
\end{figure}

The data for this study is taken from the National Traffic Information Service (NTIS)\footnote{Technical details of the NTIS data feeds are available at \url{http://www.trafficengland.com/services-info}}.
NTIS provides both historic and real-time traffic data for all roads in the UK that lie on the `Strategic Road Network.'
This network includes all motorways and major A-roads in the UK but we choose to focus our analysis on one the countries busiest motorways, the M25 London Orbital, pictured in Fig. \ref{fig:M25Plot}.
Inside NTIS, roads are represented by a directed graph, with edges (referred to as links from now on) being segments of road that have a constant number of lanes and no slip roads joining or leaving. 
We extract all links that lie on the M25 in the clockwise direction, yielding a subset of the road network to collect data for.
Placed along links are physical sensors called `loops', which record passing vehicles and report averaged quantities each minute.

The most relevant components of NTIS to our work are event flags that are manually entered by traffic operators.
These flags specify an event type, for example accident or obstruction, the start and end time and the link the event occurred on.
We extract all accidents and obstructions that occurred on the chosen links between September 1st 2017 and September 30th 2018.
However, we require more fine-grained locations for events than just the link it occurred on, so we perform further localisation.
This pre-processing step is highly specific to our dataset, so we only offer an overview here, with more details given in the supplementary material.  
To understand why this is needed, one should note that links vary in size from around 500 to 10,000 meters. 
However, the average distance between successive loop sensors is roughly 500 meters.
As a result, the data is not initially appropriate to model with point-processes, however one can use the time-series provided by individual loop sensors to determine higher resolution location data for events.
Given an event flag and link, we localize the event in-between the two loop sensors on that link that show the largest drop in speed and rise in occupancy when going from the sensor downstream to upstream. 
This is motivated by existing work in incident detection, with an example being \cite{california_algorithm_original}.

\section{Methodology}\label{sec:Methodology}

\subsection{Model formulation}

Our objective is to model the number of events (incidents) observed between any two positions $x_1$ and $x_2$ in any time interval between $t_1$ and $t_2$. 
We denote this quantity by $N_{[x_1,x_2), [t_1, t_2)}$.
Since we know that accidents cluster in both space and time, the simplest model is a non-homogeneous Poisson point process. 
This model is specified by an underlying intensity function, $\lambda(x,t)$, which is the local event probability per unit length per unit time.  
It is then assumed that for any intervals, $[x_1, x_2)$ and $[t_1, t_2)$, $N_{[x_1,x_2), [t_1, t_2)}$ has a Poisson distribution:
\begin{equation}
\mathbb{P}( N_{[x_1,x_2), [t_1, t_2)}= n) = \frac{\Lambda_{[x_1,x_2), [t_1, t_2)}^n}{n !} \exp\left( - \Lambda_{[x_1,x_2), [t_1, t_2)}\right),
\end{equation}
where
\begin{equation}
\Lambda_{[x_1,x_2), [t_1, t_2)} = \int_{t_1}^{t_2}\int_{x_1}^{x_2} \lambda(x, t) dx\,dt. 
\end{equation}

It is natural to assume that the intensity is multiplicatively decomposable:
\begin{equation}\label{equ:intensity}
\lambda(t,x) = \mu_0\,\mu_d(t)\,\mu_w(t)\,\mu_t(t)\,\mu_s(x).
\end{equation}
Here $\mu_0$ is a uniform background intensity which has units of events per unit time per unit length.
This uniform background is then modulated by the functions $\mu_d(t)$, $\mu_w(t)$,  $\mu_t(t)$ and $\mu_s(x)$ to capture spatio-temporal variation.
The spatial modulation, $\mu_s(x)$,  accounts for the fact that  different locations will naturally have differing rates of accidents. 
For example junctions with low visibility having higher rates than straight, simple sections of road.
The temporal modulation, $\mu_d(t)\,\mu_w(t)\,\mu_t(t)$ consists of three components: $\mu_d(t)$ represents daily variation, $\mu_w(t)$ represents weekly variation and $\mu_t(t)$ represents any long term trend that may be present.
Daily and weekly seasonality is a ubiquitous feature of traffic data reflecting daily `rush-hour' commuting patterns, and weekly differences between the 5-day working week to the 2-day weekend. 
Annual seasonality may also be present but since our data spans one year, any such variation is captured by the trend.

Inhomogeneous point processes describe clustering solely in terms of variations in the intensity function. 
A more sophisticated model is a self-exciting point process, known as a Hawkes process, to capture the distinction between primary and secondary incidents. 
Self-excitation means that when an event occurs, the probability of observing a subsequent event nearby increases.
A second term depending on the previous events is added to Eq.~(\ref{equ:intensity}) to give what is called the {\em conditional} intensity function:
\begin{equation}\label{equ:BasicModelEquation}
\lambda(t,x) = \mu_0\,\mu_d(t)\,\mu_w(t)\,\mu_t(t)\,\mu_s(x) + A\sum_{\substack{t_i < t \\ x_i > x}}\,g(t-t_i)h(x-x_i),
\end{equation}
where $A$ is the triggering rate, $g$ and $h$ are triggering functions that describe how the triggering mechanism decays in time and space respectively and $(t_i, x_i)$ are the times and locations of the observed incidents.
The word `conditional' here reflects the dependence of the intensity on the realisation of the process.

Our methodology is based on the work of \cite{a_semiparametric_spatiotemporal_Hawkes_type_point_process_model_with_periodic_background_for_crime_data}, where it is shown how to construct the conditional intensity of a Hawkes process in a non-parametric way by applying kernel smoothing to the observed data on crime.
We modify this model to make it applicable to traffic data.
The supplementary material contains the model derivation, estimators of model components and details of the fitting algorithm, following closely  \cite{a_semiparametric_spatiotemporal_Hawkes_type_point_process_model_with_periodic_background_for_crime_data}.
There are three main changes. 
Firstly, the spatial triggering mechanism is one-dimensional and unidirectional: secondary incidents cannot occur downstream of the primary incident.
Secondly we enforce monotonicity of the triggering functions, $g$ and $h$, in Eq.~(\ref{equ:BasicModelEquation}) as a constraint to help with identifiability. 
Thirdly, we apply boundary correction to the kernel density estimates to reduce bias.
In the remainder of this section, further detail is provided on these changes.

\subsection{Reconstructing One-Dimensional and Unidirectional Spatial Triggering}

First, we adapt the work in \cite{a_semiparametric_spatiotemporal_Hawkes_type_point_process_model_with_periodic_background_for_crime_data} to reconstruct a one-dimensional and unidirectional spatial triggering function. 
To begin, we state that for a spatio-temporal point process $N_p$ with conditional intensity function $\lambda(t,x)$ and a predictable process $f(t,x) \geq 0$, over a time domain $[T_1, T_2]$ and space domain $S$, we can write:
\begin{equation}\label{equ:ExpetationTheorem}
\mathbb{E}\left[ \int_{[T_1, T_2] \times S} f(t,x)dN_p(dt \times dx) \right] = \mathbb{E} \left[ \int_{T_1}^{T_2} \int_{S} f(t,x)\lambda(t,x) dxdt \right]. 
\end{equation}
To determine our triggering functions in an entirely data-driven way, we first consider two data points $\left(\tau^{(1)}, \chi^{(1)}\right)$ and $\left(\tau^{(2)}, \chi^{(2)}\right)$ where $\tau^{(1)} < \tau^{(2)}$ and $ \chi^{(1)} > \chi^{(2)}$.
We then define $\rho(\tau^{(1)}, \chi^{(1)}, \tau^{(2)}, \chi^{(2)})$ as:
\begin{equation}
\rho\left(\tau^{(1)}, \chi^{(1)}, \tau^{(2)}, \chi^{(2)}\right) = \begin{cases}
      \frac{A g( \tau^{(2)} - \tau^{(1)} )h( \chi^{(2)} - \chi^{(1)} )}{\lambda(\tau^{(2)}, \chi^{(2)})}, & \text{if}\ \tau^{(1)} < \tau^{(2)}  \text{and}\ \chi^{(1)} > \chi^{(2)}. \\
      0, & \text{otherwise}
    \end{cases}
\end{equation}
If we apply this in Eq.~(\ref{equ:ExpetationTheorem}), letting $f(\tau, \chi) = \rho(t_i, x_i, \tau, \chi)\mathbbm{1}_{ \chi - x_i \in [x - \Delta x, x + \Delta x] }$ for some small positive $\Delta x$, on the temporal domain $[0,T]$ and spatial domain $[0,X]$, we attain:
\begin{equation}\label{equ:DeriveTemporalTrigger}
\begin{split}
&\sum_{j}\rho(t_i, x_i, \tau_j, \chi_j)\mathbbm{1}_{ \chi - x_i \in [x - \Delta x, x + \Delta x] } \\
&\approx \int_{0}^T \int_{0}^X \rho(t_i, x_i, \tau, \chi)\mathbbm{1}_{ \chi - x_i \in [x - \Delta x, x + \Delta x] } \lambda(\tau, \chi) d\chi d\tau \\
&= A\int_{t_i}^T \int_{0}^{x_i}  \frac{g(\tau - t_i) h(\chi - x_i)}{\lambda(\tau, \chi)} \mathbbm{1}_{ \chi - x_i \in [x - \Delta x, x + \Delta x] } \lambda(\tau, \chi) d\chi d\tau \\
&= A\left(\int_{t_i}^T g(\tau - t_i)d\tau\right) \left( \int_{0}^{x_i} h(\chi - x_i)\mathbbm{1}_{ \chi - x_i \in [x - \Delta x, x + \Delta x] } d\chi \right).
\end{split}
\end{equation} 
If we then let $s = \chi - x_i$, we have:
\begin{equation}
\begin{split}
&\sum_{j}\rho(t_i, x_i, \tau_j, \chi_j)\mathbbm{1}_{ \chi_j - x_i \in [x - \Delta x, x + \Delta x] } \\
&\approx A\left(\int_{t_i}^{T} g(\tau - t_i)d\tau\right) \left( \int_{-x_i}^{0} h(s)\mathbbm{1}_{ s \in [x - \Delta x, x + \Delta x] } ds \right) \\
&= A\left(\int_{t_i}^{T} g(\tau - t_i)d\tau\right) \left( \int_{x-\Delta x}^{x+\Delta x} h(s) ds \right) \\ 
&\propto h(x). 
\end{split}
\end{equation}
and hence:
\begin{equation}\label{equ:TemporalTriggerHistogram}
\hat{h}(x) \propto \sum_{(i, j)}  \rho_{i,j}\mathbbm{1}_{ x_j - x_i \in [x - \Delta x, x + \Delta x] }
\end{equation}
with:
\begin{equation}\label{equ:rhoij}
\rho_{i,j} = \frac{Ag(t_j-t_i)h(x_j-x_i)}{\lambda(t_j,x_j)}, \, \, \text{for all} \, \, (i,j) \ \text{s.t.}\ t_i < t_j \ \text{and}\ x_i > x_j. 
\end{equation}

This derivation follows closely that in \cite{a_semiparametric_spatiotemporal_Hawkes_type_point_process_model_with_periodic_background_for_crime_data}, with an additional restriction on $\rho_{i,j}$, and $h$ being a univariate function.
One then smooths this initial estimate using kernel functions, with more details on this and all other model components in the supplementary material.

\subsection{Constraining Triggering Functions}

We note that this model aims to explain any residuals from the background process using the triggering component.
As such, the model has significant freedom to adapt to the data. 
We can limit this freedom to ensure that the triggering component truly reflects increased rates of events on a short-time scale (compared to the periodic components) in the wake of a particular event.
A natural way to do this and extension of the original methodology is to enforce the triggering functions to be monotonic.
This is a reasonable constraint that ensures interpretability of the triggering functions, whilst still allowing them to be constructed by the data.
Triggering decaying in space and time suggests we expect reduced influence of an event as we move further from its location and as time passes.
To accomplish this, we follow the method in \cite{non_parametric_kernel_regression_subject_to_monotonicity_constraints}. 
In standard kernel smoothing, one writes a smoothed estimate of some dataset (X,Y) as:
\begin{equation}\label{equ:KernelSmoothOriginal}
\hat{\nu}(x) = \frac{1}{N}\sum_{i=1}^N \frac{1}{h}K\left(\frac{x-X_i}{h}\right)Y_i
\end{equation}
with some kernel $K$, some bandwidth $h$, and attain a smoothed estimate $\hat{\nu}(x)$.
Specifically in our application, $X$ represents differences in event locations or event times, and all $Y$ values equal 1, however the technique works for more general cases.
\cite{non_parametric_kernel_regression_subject_to_monotonicity_constraints} generalized Eq.~(\ref{equ:KernelSmoothOriginal}) to incorporate a weight $p_i$ to each data-point used in the smoothing to `adjust' the initial smoothed fit to be monotonic.
One writes this adjusted fit as: 
\begin{equation}\label{equ:KernelSmoothMono}
\hat{\nu}_{mono}(x | p_1, \dots, p_N ) = \frac{1}{N}\sum_{i=1}^N \frac{1}{h}K\left(\frac{x-X_i}{h}\right)Y_ip_i.
\end{equation}
Our goal is now to choose a weight $p_i$ for each data-point $i$ used in the construction of the function, whilst altering the original estimate as little as possible.
There are an infinite number of sets of $\{ p_1, p_2, \dots, p_N \}$ one could choose to enforce a monotonic function, and to identify a unique solution, we choose the set that is as close to the uniform distribution $\left\{ \frac{1}{N}, \dots , \frac{1}{N} \right\}$ as possible.
One distance measure used in the work to compare the $p_i$'s to a uniform distribution is:
\begin{equation}\label{equ:DistFuncMono}
D_0(p_1, \dots, p_N) = -\sum_{i=1}^N \log(Np_i). 
\end{equation}
Using this, we can then introduce a step in our model fitting where we solve a further optimization problem, determining each $p_i$ value to produce a monotonic triggering function.
The optimization problem is specified as: 
\begin{equation}\label{equ:OptimProblemMonotonic}
\begin{aligned}
\min_{p_1, \dots, p_N} \quad & D_0(p_1, \dots, p_N)  \\
\textrm{s.t.} \quad & \frac{d\hat{\nu}_{mono}(x | p_1, \dots, p_N )}{dx} \leq \epsilon \\
					& p_i \geq 0 \, \, \, \text{for all} \, \, i \in [1, 2, \dots, N] \\
					& p_i \leq 1 \, \, \, \text{for all} \, \, i \in [1, 2, \dots, N] \\
					& \sum_{i=1}^N p_i = 1.
\end{aligned}
\end{equation}
Since we are enforcing the triggering to decay with increasing time and distance, $\epsilon$ constrains the gradient of the triggering functions to be below some value.
It should be noted that in practice, constraining the triggering function only alters the resulting functions in section \ref{sec:ResultsSeasonal}, where we fit the model to small subsets of data.
When fit to the entire year, the resulting functions are already monotonic.

\subsection{Boundary Correction}

It is well known that, on a truncated domain, kernel density estimates are biased near the boundary, with discussion of this in \cite{boundary_adjusted_density_estimation_and_bandwidth_selection} and references within.
There are many ways to account for this throughout the literature, with a simple method being to `mirror' the data, supposing we have extra data-points.
If we have our `real' data-points $X_1, X_2, \dots X_N$ and wish to truncate our domain at 0, we introduce extra points $-X_1, -X_2, \dots -X_N$.
The smoothed estimate $\hat{\nu}(x)$, using a bandwidth $h$, is then:
\begin{equation}\label{equ:KernelSmoothOriginalBoundaryCorrect}
\hat{\nu}(x) = \frac{1}{Nh}\sum_{i=1}^N \left[ K\left(\frac{x-X_i}{h}\right) + K\left(\frac{x+X_i}{h}\right) \right].
\end{equation}
We apply this correction to both triggering functions and the background trend. 

Other components in our model are not truncated at a specific point, but rather periodic over some domain.
This is true for the daily and weekly background components, but we further assume that the spatial background is periodic, as the M25 is an almost continuous ring around London.
Recall Fig. \ref{fig:M25Plot}, a small section of the M25 is not a motorway and reports no data, but it is negligible in comparison to the wider motorway, so assuming a spatial background on a ring is reasonable.
These components do not strictly need boundary correction, however care must be taken in their evaluation.
Suppose we take the daily background as an example.
We observe events times $t_i$ on a domain $[0,T]$ and map these onto some periodic domain $[0, m_d]$ where $m_d$ represents the number of minutes in a day.
The mapped times would then be given by $t_i - m_d\left\lfloor \frac{t_i}{m_d} \right\rfloor$. 
However, an event may have a non-negligible contribution from its location mapped to the previous and following day, that is we observe a contribution from times $t_i - m_d\left\lfloor \frac{t_i}{m_d} \right\rfloor - m_d$ and $t_i - m_d\left\lfloor \frac{t_i}{m_d} \right\rfloor + m_d$. 
As a result, the periodic estimates are the sum over all data-points of the kernel evaluated both inside the domain, and at one preceding and following period.
This could actually be an infinite sum, over the entire past and future days, however in practice the decay of a Gaussian kernel leads to it being sensible to truncate this sum to just these terms.
More details of this are provided in the supplementary material.

\section{Results}\label{sec:Results1Year}

\subsection{Bandwidth Selection, Model Selection \& Prevalence of Triggering}\label{subsec:ResultsModelSelection}

Selection of smoothing bandwidths is an open problem for models of this form, however the application discussed offers natural choices with time-scales inherent to the system.
We choose the daily, weekly and trend bandwidths to be $60$, $10\times60$ and $60\times24\times14$ minutes respectively.
The choice of daily bandwidth is selected due to the `rush-hour' behaviour in the UK typically varying on a time-scale of around an hour, whilst the weekly and trend components capture variation across larger time-scales.
The spatial bandwidth is chosen as $5500$ meters, which appears small enough to capture differing features across the M25, whilst not introducing superfluous oscillations.
This is also larger than the uncertainty we would expect in event locations, therefore accounting for potential uncertainty in the data. 
Finally, the temporal and spatial bandwidths for the triggering functions are chosen to be $30$ minutes and $500$ meters respectively. 
We considered variations on all of these values, finding those listed provided a reasonable compromise of model interpretability and identify known components of traffic flow.   

We first consider some measure of goodness of fit for models containing variations of the discussed components. 
The first is a homogeneous Poisson process, used as the simplest reference model one could construct.
We then compare models with: daily and weekly background components, daily, weekly and trend background components, daily, weekly background components and triggering, and daily, weekly, trend background components and triggering. 
To compare these, we consider the log-likelihood, given by:
\begin{equation}\label{equ:LogLikelihoodFirst}
\log(L) = \sum_{i=1}^{N} \log\left( \lambda(t_i, x_i) \right) - \int_{0}^T\int_{0}^X \lambda(t,x) dxdt,
\end{equation}
with a larger value suggesting a better model.
Results for this are given in table \ref{table:LogLikelihoods}.
Note that in using log-likelihood to judge the goodness of fit, we ignore model complexity, the idea that we could continuously add components to any model and see increasingly marginal improvements as we attain a more complex model. 
However, our results throughout this section show the estimated functions present reasonable and interpretable behaviour that seems to catch marginal but still present features of the phenomenon of interest. To check the specification of the models we inspect the residuals and validate the results in section \ref{sec:ModelValidationFullData}, while a successful attempt of out-of-sample validation is attained in section \ref{sec:ResultsSeasonal}.
\begin{table}
\caption{ Log-likelihood values for models with various components.\label{table:LogLikelihoods}}
\centering
\fbox{%
\begin{tabular}[ht]{|c|c|c|}
\hline
Model                               & $A$       & Log-Likelihood \\
\hline
Fixed Rate Poisson Process          & -         & -28861.55  \\
\hline
Daily + Weekly Background           & -         & -28028.05  \\
\hline
Daily + Weekly  + Trend Background  & -         & -27929.60  \\
\hline
Daily + Weekly  + Triggering        & 0.068495  & -27864.48  \\
\hline
Daily + Weekly + Trend + Triggering & 0.065462  & -27781.38  \\
\hline
\end{tabular}}
\end{table}

Inspecting table \ref{table:LogLikelihoods}, we see by far the worst model is a constant rate Poisson process, and adding periodic daily and weekly components to this shows the largest improvement in log-likelihood. Including a trend and triggering component further improve the log-likelihood.
Using this methodology, the parameter $A$ can be interpreted as the proportion of the impact of the triggering function on the total intensity.
For our optimal model incorporating all components, this means about 6.55\% of events appear to be the result of triggering, in practical terms about 100 events.
One may want to consider this as an upper bound, as discussed in section \ref{sec:ResultsTriggering}, along with an appropriate time-scale.

\subsection{Background Analysis}\label{sec:ResultsBackground}

\begin{figure}[!ht]
     \centering
     \subfloat[][Daily Background]{\includegraphics[width=0.31\textwidth]{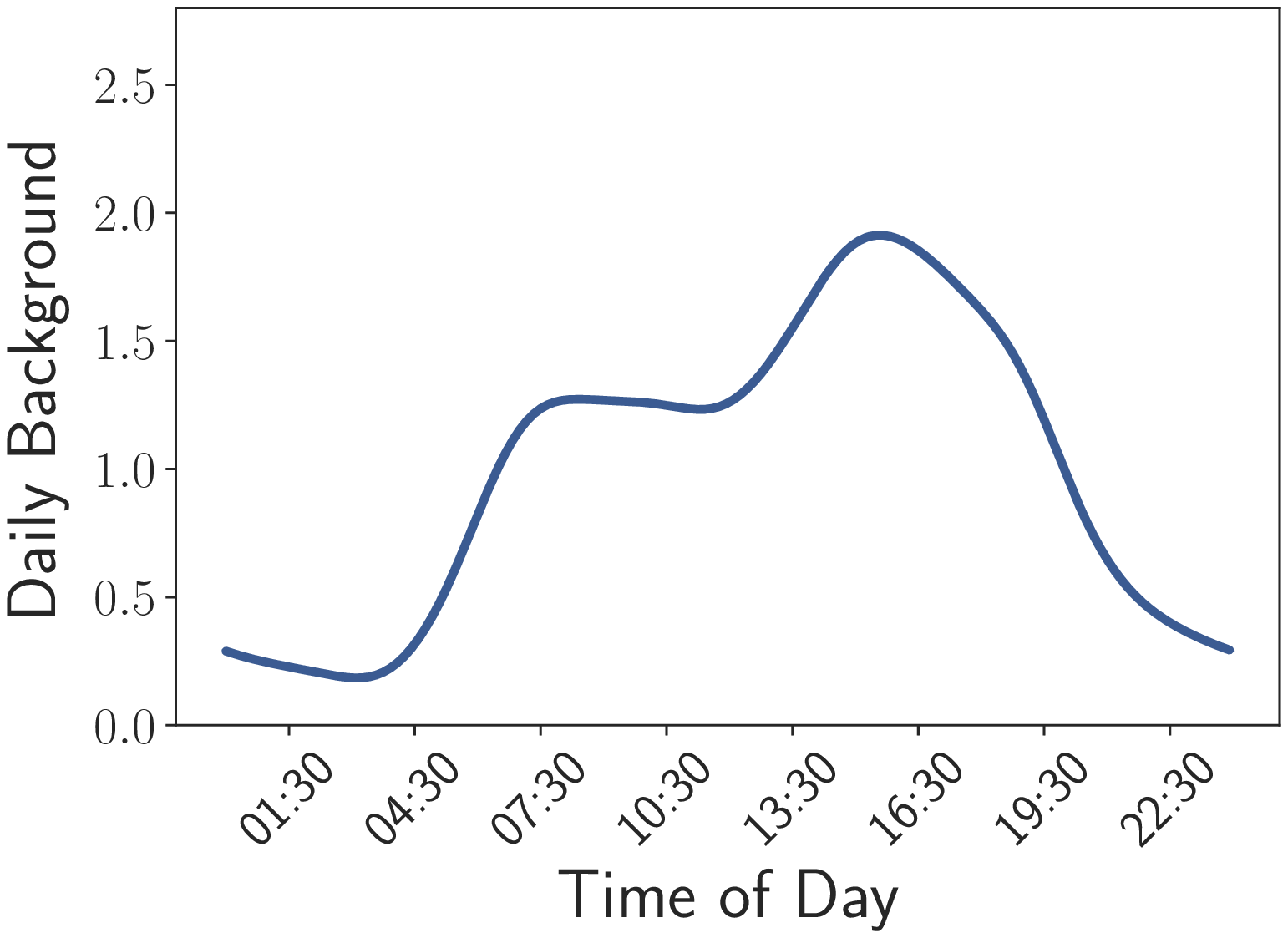}\label{fig:DailyBackground12Months}}
     \subfloat[][Weekly Background]{\includegraphics[width=0.31\textwidth]{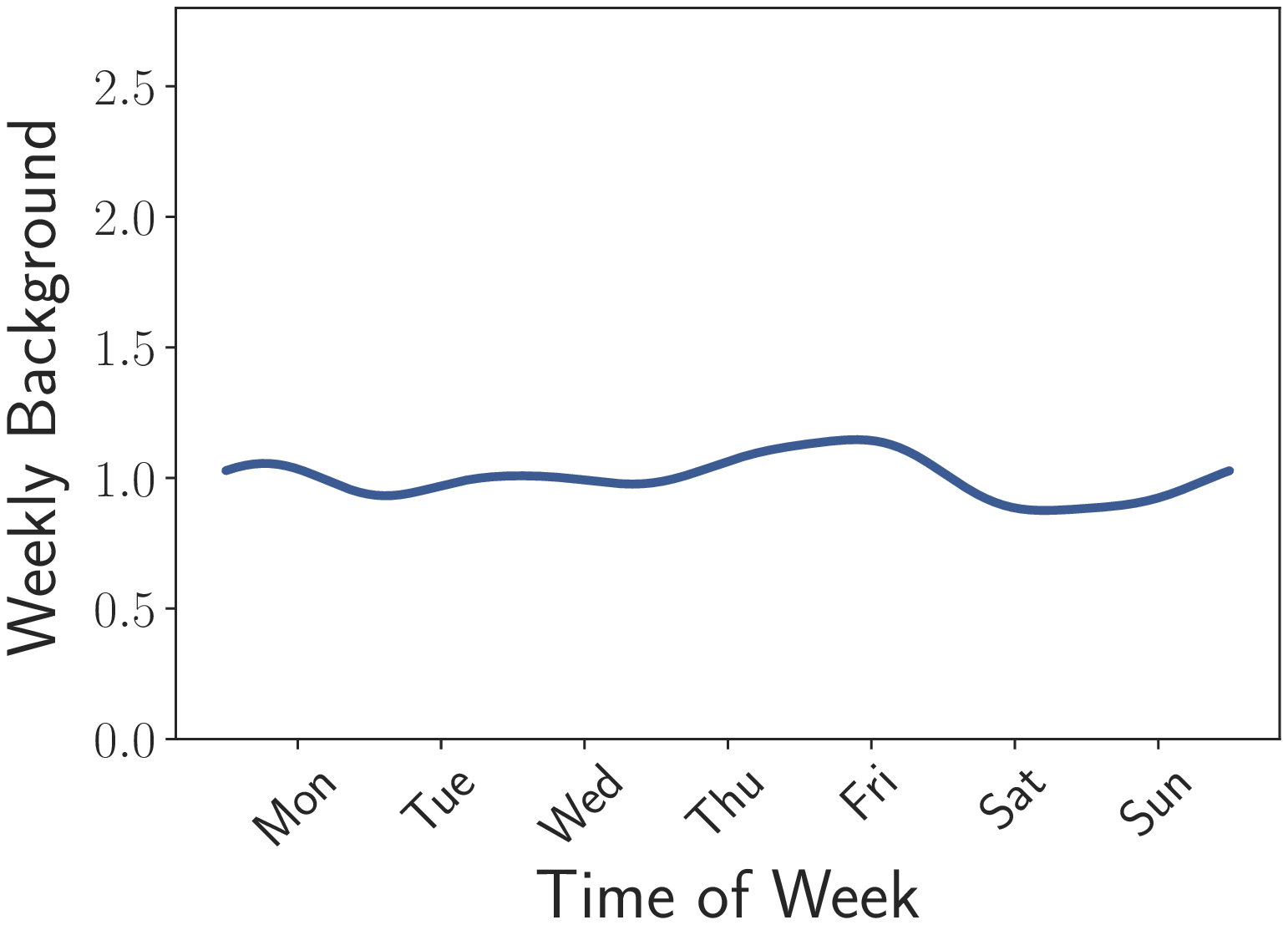}\label{fig:WeeklyBackground12Months}}
     \subfloat[][Trend Background]{\includegraphics[width=0.31\textwidth]{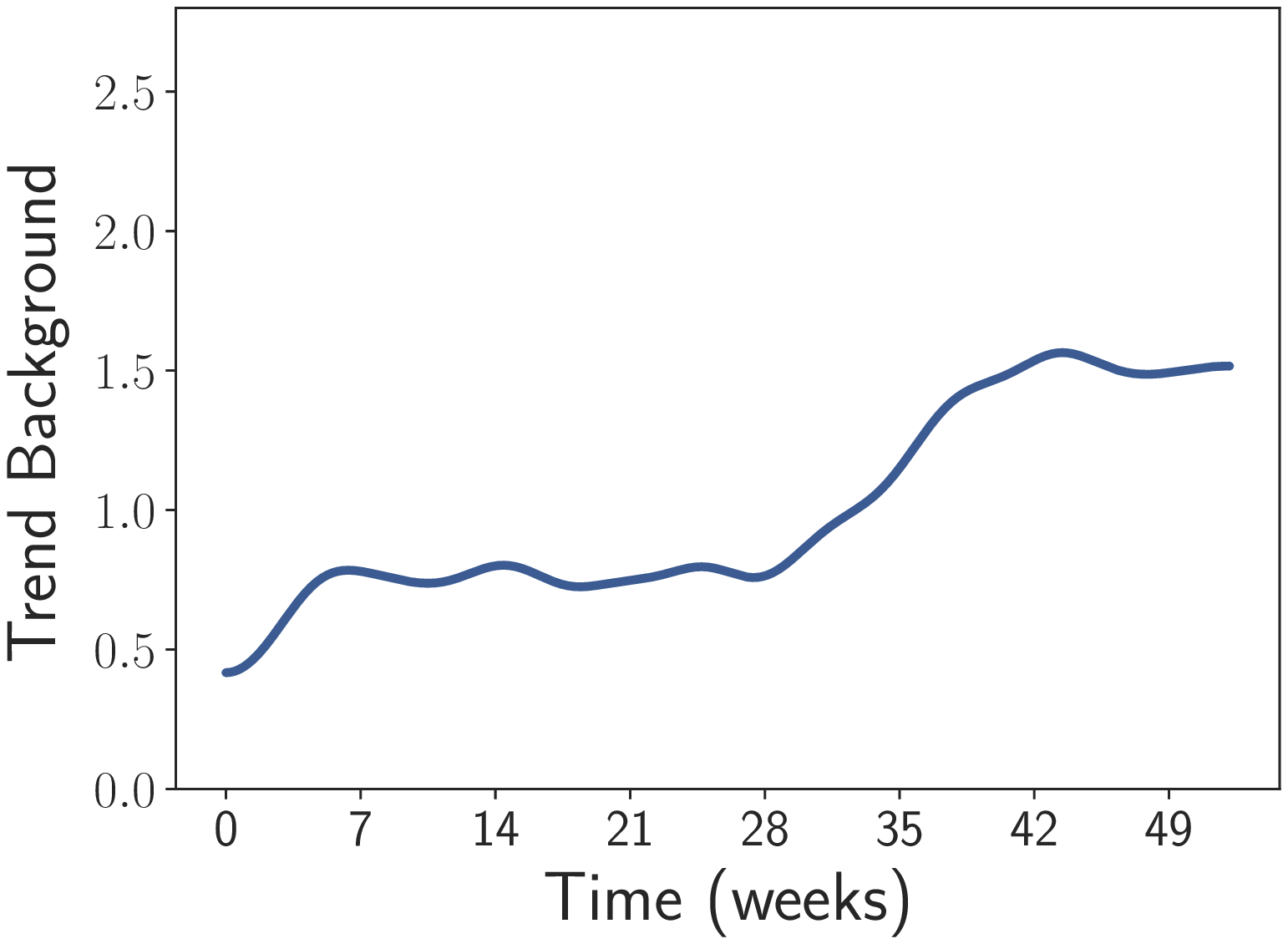}\label{fig:SpatiaTrendBackground12Months}}
	 \caption[Temporal Background Components - 1 Year of Data]{Temporal Background Components fit to 1 year of data. }\label{fig:TemporalBackgroundComponents1Year}

     \subfloat[][Spatial Background]{\includegraphics[width=0.31\textwidth]{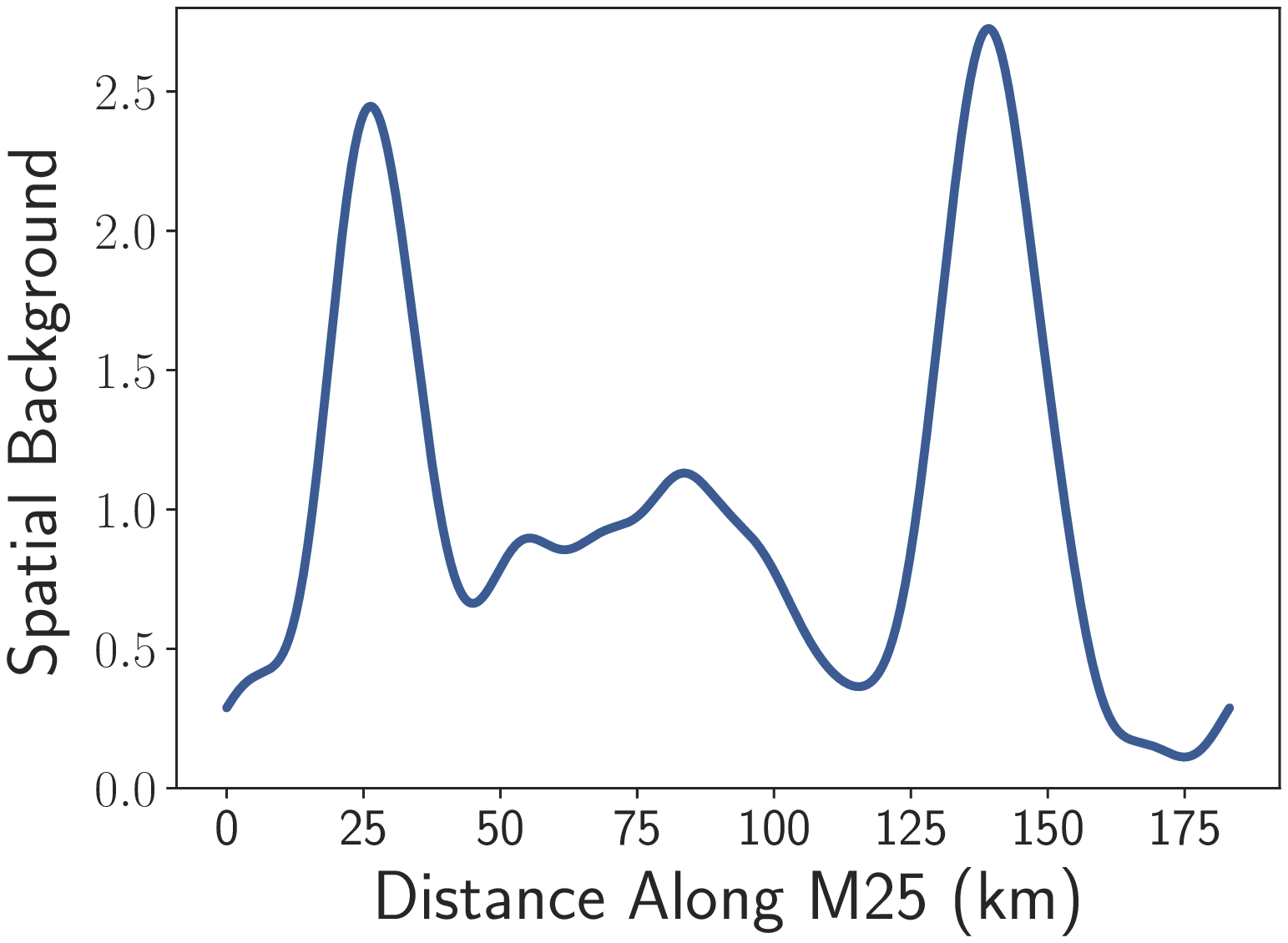}\label{fig:SpatialBackground12Months}}
     \subfloat[][Spatial Background overlaid on a map of the M25.]{\includegraphics[width=0.31\textwidth]{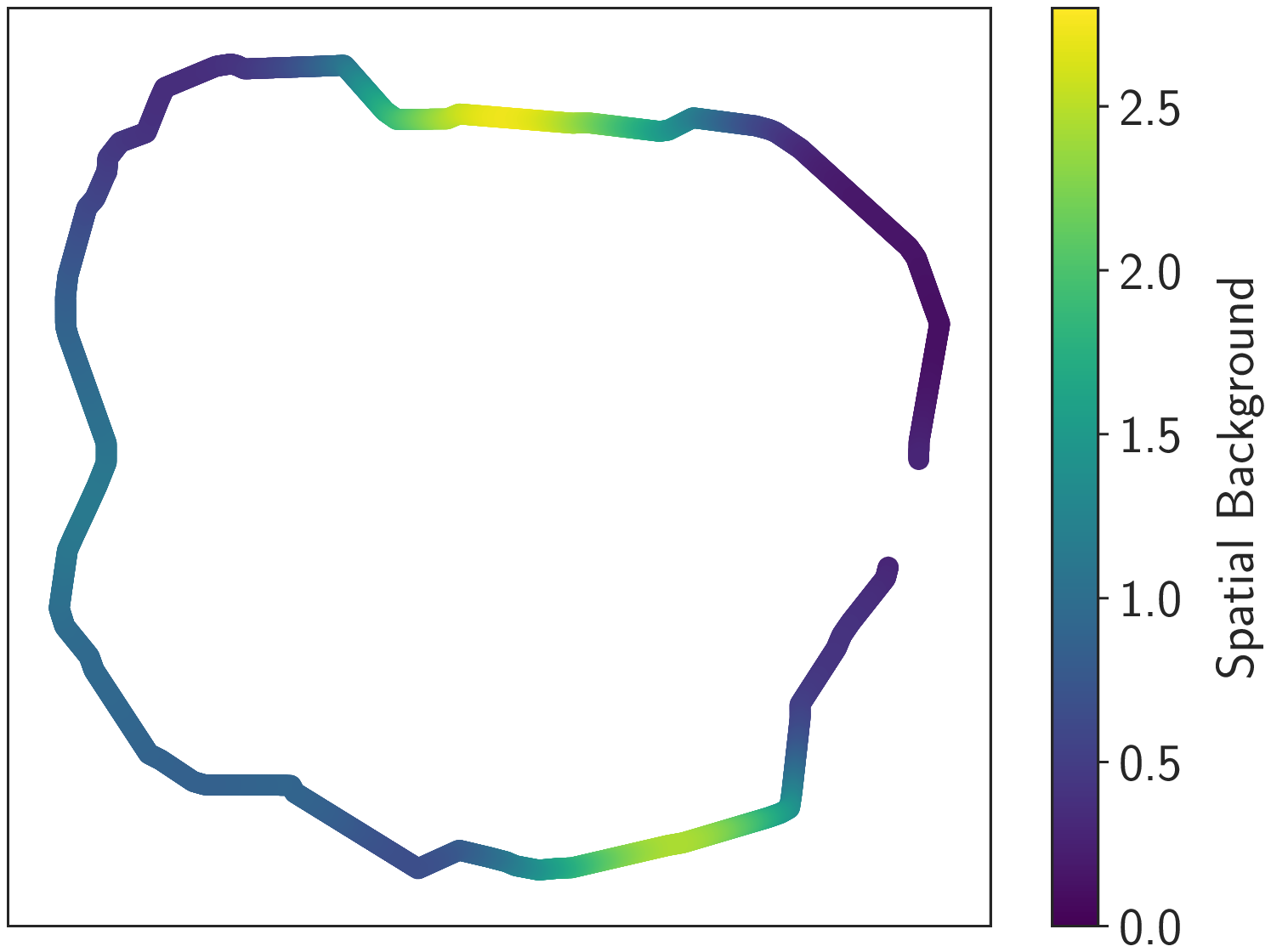}\label{fig:MapBackground12Months}}
     \caption{Spatial Background Components fit to 1 year of data. }\label{fig:SpatialBackgroundComponents1Year}
\end{figure}

The background component of the model is strong, as seen when inspecting the changes in log-likelihood from table \ref{table:LogLikelihoods}.
In our model, this constitutes a daily, weekly, spatial and trend component.
We visualize the temporal background components in Fig. \ref{fig:TemporalBackgroundComponents1Year}.

Inspecting Fig. \ref{fig:DailyBackground12Months}, we see that the daily background increases to an initial peak during the morning rush hour, then remains roughly constant, before rising again to a peak at around 4pm, and decaying after this. 
From Fig. \ref{fig:WeeklyBackground12Months}, there appears to be much less variation in the intensity across the week compared to all other identified components, but a slightly higher intensity on Thursdays and Fridays, and the lowest on Tuesdays and Saturdays. 
Finally, we see a small increase in the trend during the first 7 weeks of the data, then it remains reasonably flat until week 28, where it begins to rise again. Around week 40, it stabilizes again.
This could be due to an increase in actual event intensity, or more comprehensive reporting after a certain point, more operators and so forth, however no changes in reporting are known to us.

As-well as the temporal background, where events are most common around the M25 is of interest. 
We show the spatial background in Fig. \ref{fig:SpatialBackgroundComponents1Year}, from which a clear spatial structure is visible.
We see two distinct peaks in Fig. \ref{fig:SpatialBackground12Months}, and a smaller spread out peak in-between these two.
The largest peak, around 140 kilometres along the motorway, is located near the `Potters Bar' junction.
The second largest, around 25 kilometres into the motorway, is located between where the M25 meets the M26, and where the M25 meets the M23. 
This background intensity, imposed onto a map schematic of the real M25 is plotted in Fig. \ref{fig:MapBackground12Months}.
We investigated if the temporal background in the vicinity of these spatial peaks differed to that across the entire M25, but found only minor changes.
Interested readers can find this analysis in the supplementary material.

\subsection{Triggering Analysis}\label{sec:ResultsTriggering}

\begin{figure}[!ht]
     \centering
     \subfloat[][Temporal Triggering]{\includegraphics[width=0.32\textwidth]{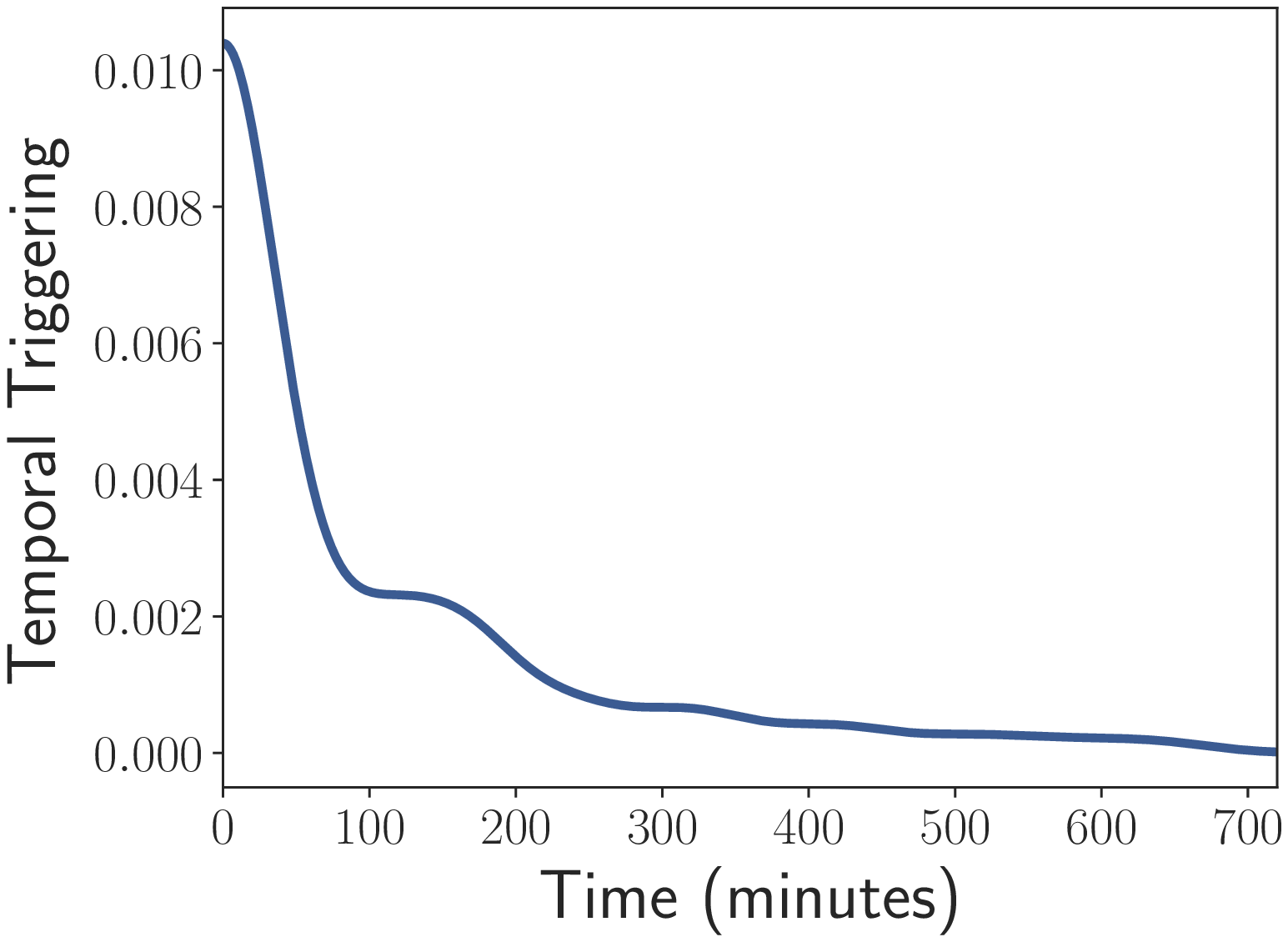}\label{fig:TemporalTrigger12Months}}
     \subfloat[][Spatial Triggering]{\includegraphics[width=0.32\textwidth]{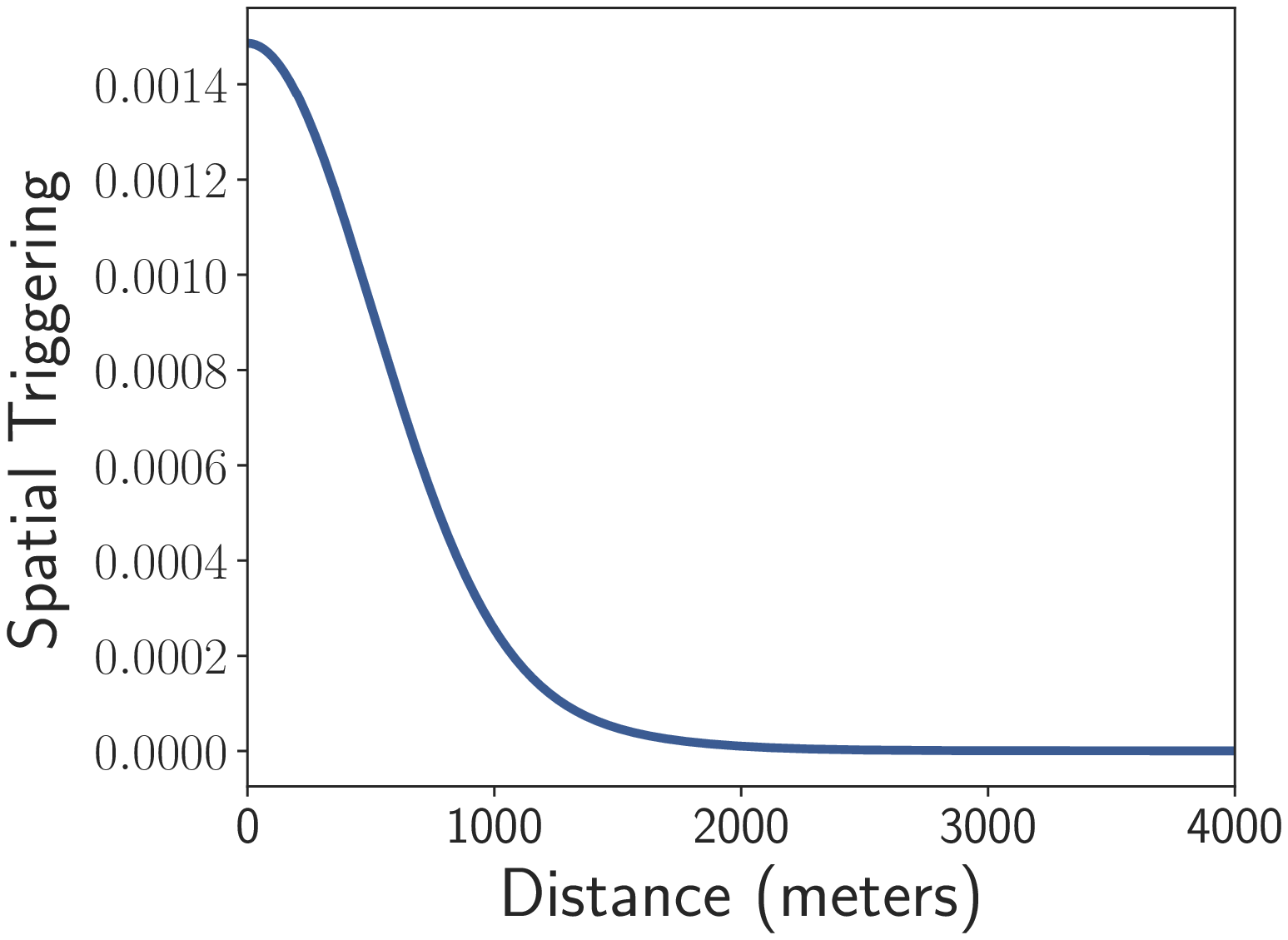}\label{fig:SpatialTrigger12Months}}
     \caption{Triggering functions, fit to 1 year of data.}\label{fig:Triggering1Year}
\end{figure}

Triggering does appear to improve the log-likelihood of our model, and as we saw in table \ref{table:LogLikelihoods} it explains around 6.55\% of events in the data. 
We visualize the resulting functions in Fig. \ref{fig:Triggering1Year}.
From Fig. \ref{fig:SpatialTrigger12Months}, it is clear that spatial triggering is limited to around 2-kilometres, after which we do not see any non-zero values.
However, temporal triggering, pictured in Fig. \ref{fig:TemporalTrigger12Months}, appears to have quite a `long tail' in the sense that it decays over a very long range.
However, this is a clear time-scale in this result of around 100 minutes.
As a result of this, one should take 6.55\% as an upper bounds of sorts, understanding this, combined with the identified time and length scales is an informative conclusion to draw.

\subsection{Model Validation}\label{sec:ModelValidationFullData}

\begin{figure}[!ht]
     \centering
     \subfloat[][CDF Validation]{\includegraphics[width=0.48\textwidth]{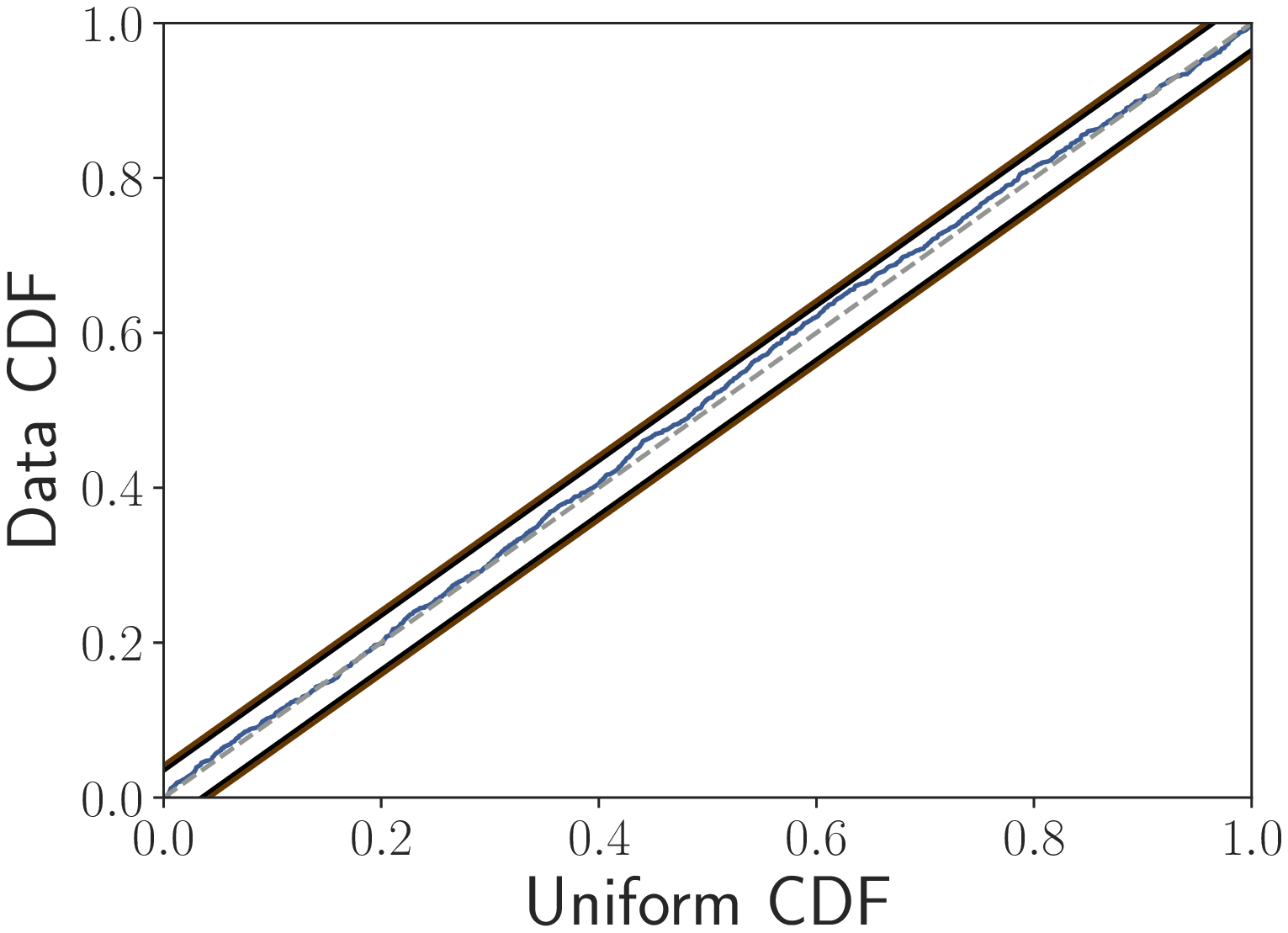}\label{fig:Dataset1CDF12Months}}
     \subfloat[][Quantile-Quantile Validation]{\includegraphics[width=0.48\textwidth]{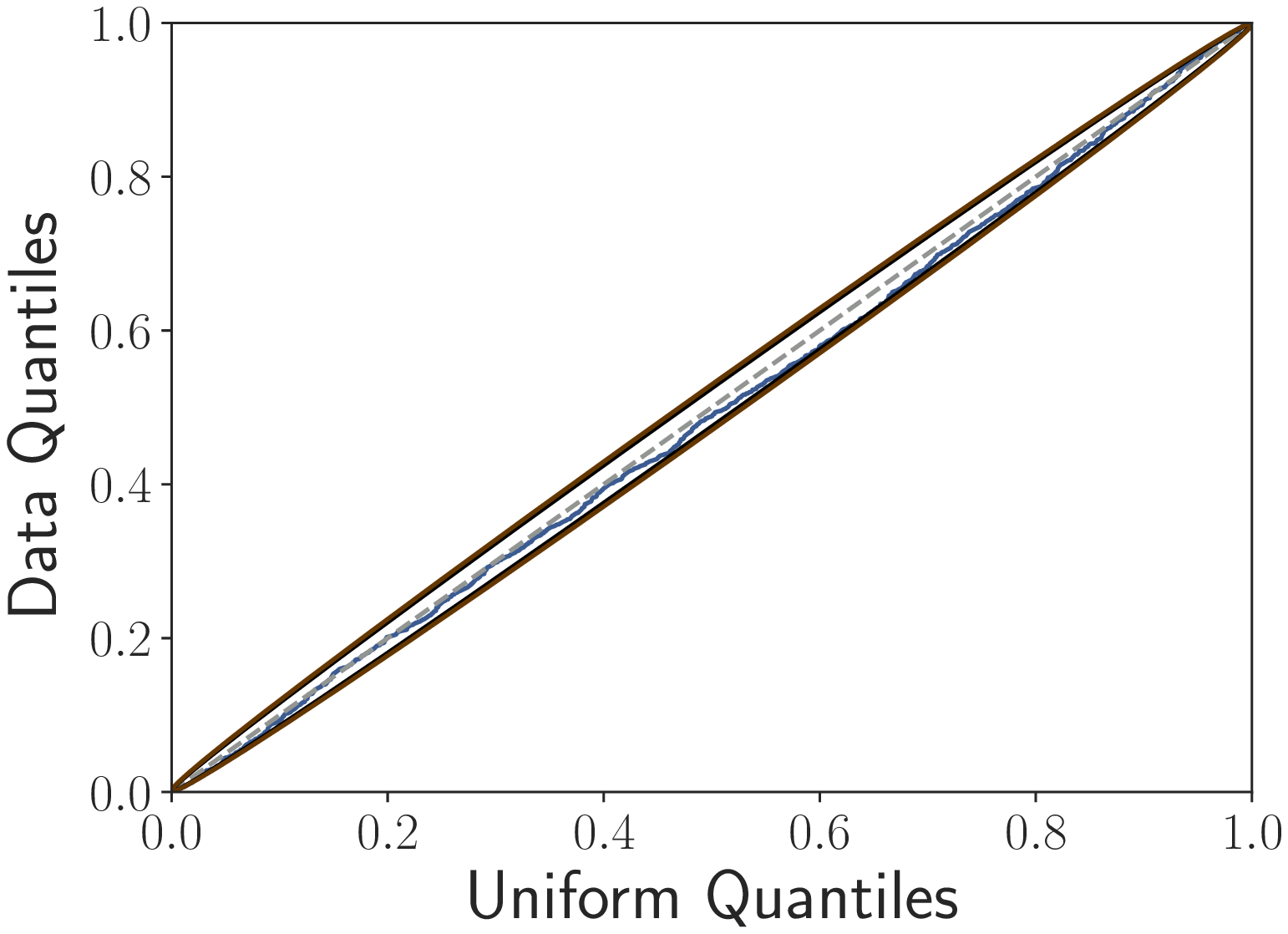}\label{fig:Dataset1QQ12Months}}
     \caption{Validation results for 1 year of data. Model results are shown with {{\DenimBlueFull}}, 95\% limits: {{\BlackFull}}, 99\% limits: {{\BrownFull}} and a reference line: {{\GreyDashed}} (colour online). }\label{fig:Dataset1ValidationCombined}
\end{figure}

To validate if our model captures the relevant features of a process, one often follows the work of \cite{statistical_models_for_earthquake_occurances_and_residual_analysis_for_point_processes}, \cite{the_time_rescaling_theorem_and_its_application_to_neural_spike_train_data_analysis} and \cite{second_order_residual_analysis_of_spatiotemporal_point_processes_and_applications_in_model_evaluation}.
Specifically, we use the transformed time-sequence given by: 
\begin{equation}\label{equ:TransformedTimeSequence}
t_i \to \Lambda_i = \int_{0}^{t_i}\int_{0}^X \lambda(u, x) dxdu,
\end{equation}
then the resulting sequence of values $\Lambda_i \, \, \text{for all} \, \, i \in \{1, 2, \dots N\}$ will follow a unit rate Poisson process if the model is correctly specified.
To prove this, one can derive that the sequence of values $\Lambda_i - \Lambda_{i-1}$ are i.i.d. random variables that follow an exponential distribution with parameter 1, which is done in \cite{the_time_rescaling_theorem_and_its_application_to_neural_spike_train_data_analysis}.
Given that we know the expected distribution of $\Lambda_i - \Lambda_{i-1}$, we can then transform this to follow a standard uniform distribution by computing: 
\begin{equation}\label{equ:ExpToUnif}
z_i = 1 - e^{-\left( \Lambda_i - \Lambda_{i-1} \right)}.
\end{equation}
Now, the computed $z_i$ values should follow the simplest known distribution, which we can evaluate by comparing the measured and theoretical cumulative distribution functions (CDF), as-well as the measured and expected quantiles.
One can generate confidence bounds for the comparison of two CDFs by inversion of the Kolmogorov-Smirnov statistic, and for quantile-quantile (QQ) plots by using the fact that the order statistics of a uniform distribution follow a beta distribution.
The distribution of the $k$-th order statistic has parameters $k$ and $n+1-k$, with $n$ being the number of sample points.
Using this, we then generate CDF and QQ plots for our different modelling scenarios, given in Fig. \ref{fig:Dataset1CDF12Months} and \ref{fig:Dataset1QQ12Months}.
It is clear from Fig. \ref{fig:Dataset1ValidationCombined} that the model is statistically defensible when inspecting both the CDF and QQ plots of the results.
Some of the quantiles in Fig. \ref{fig:Dataset1QQ12Months} are just on the edge of acceptable, but do not deviate outside of the confidence bands.
These results show that our model is well specified, however extra components could continue to be added to improve the fit. 

\subsection{Do Components Change with Season?}\label{sec:ResultsSeasonal}

\begin{figure}[!ht]
     \centering
     \subfloat[][Daily Background]{\includegraphics[width=0.31\textwidth]{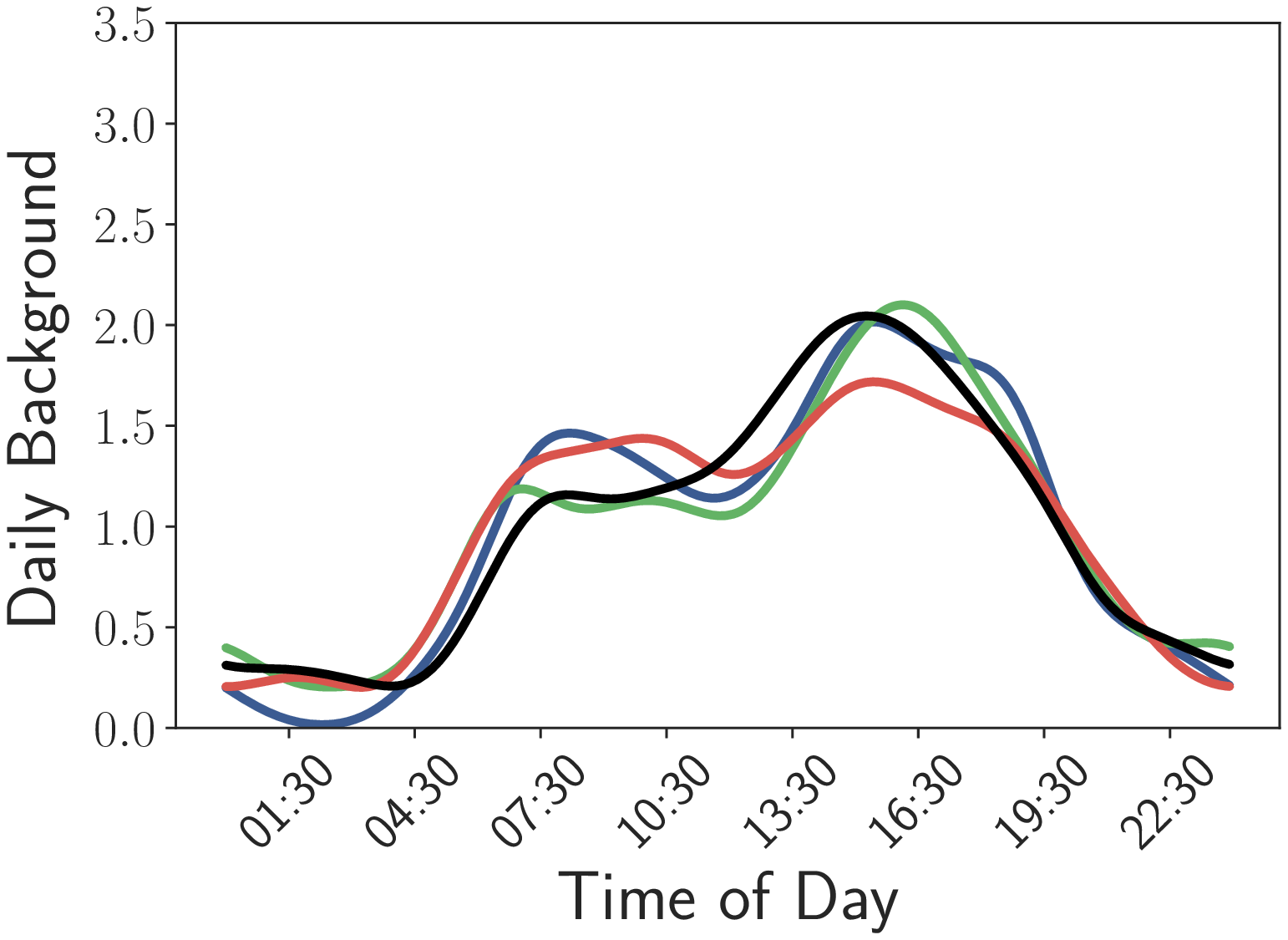}\label{fig:DailyBackground3Months}}
     \subfloat[][Weekly Background]{\includegraphics[width=0.31\textwidth]{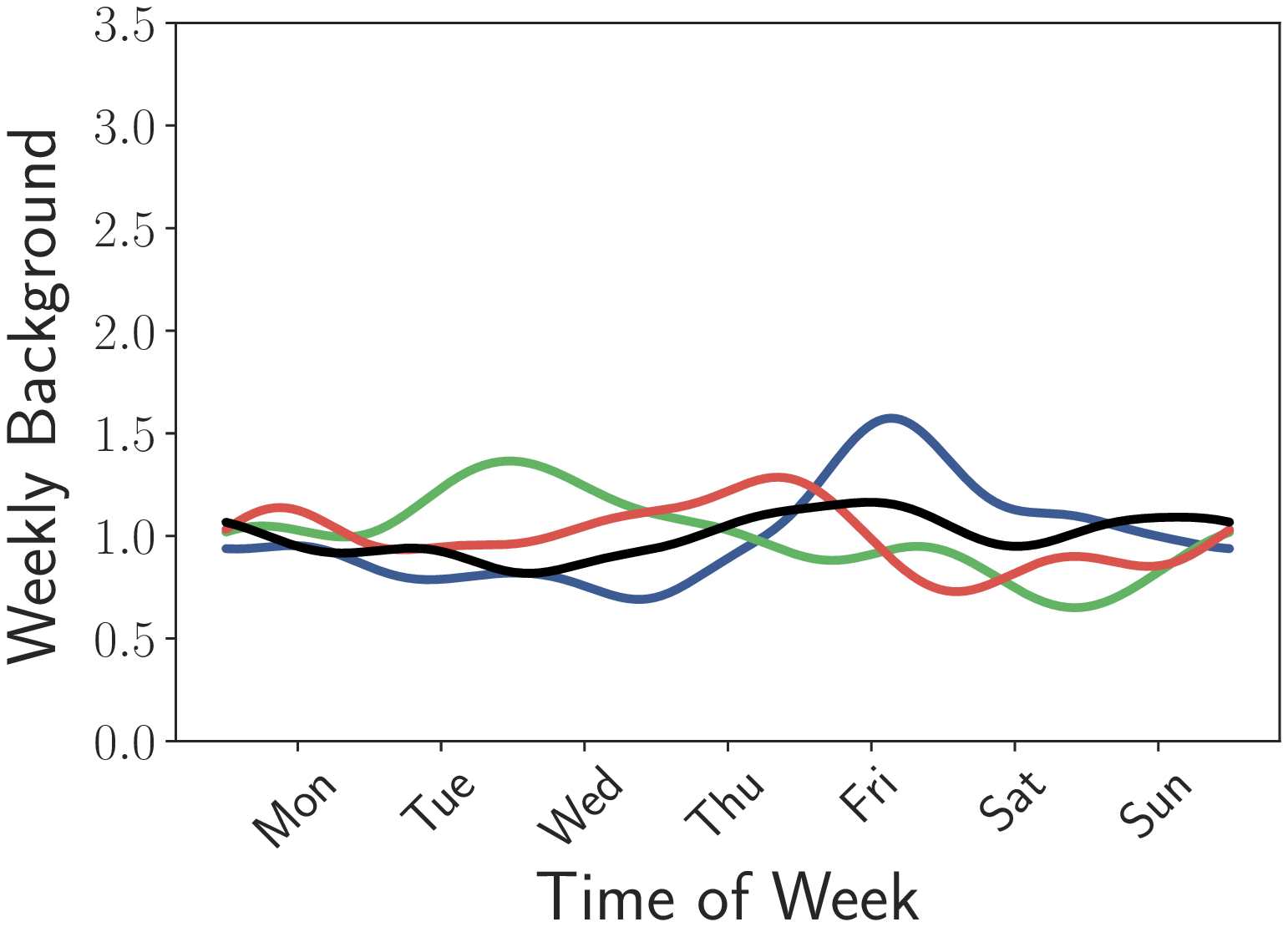}\label{fig:WeeklyBackground3Months}}
     \subfloat[][Spatial Background]{\includegraphics[width=0.31\textwidth]{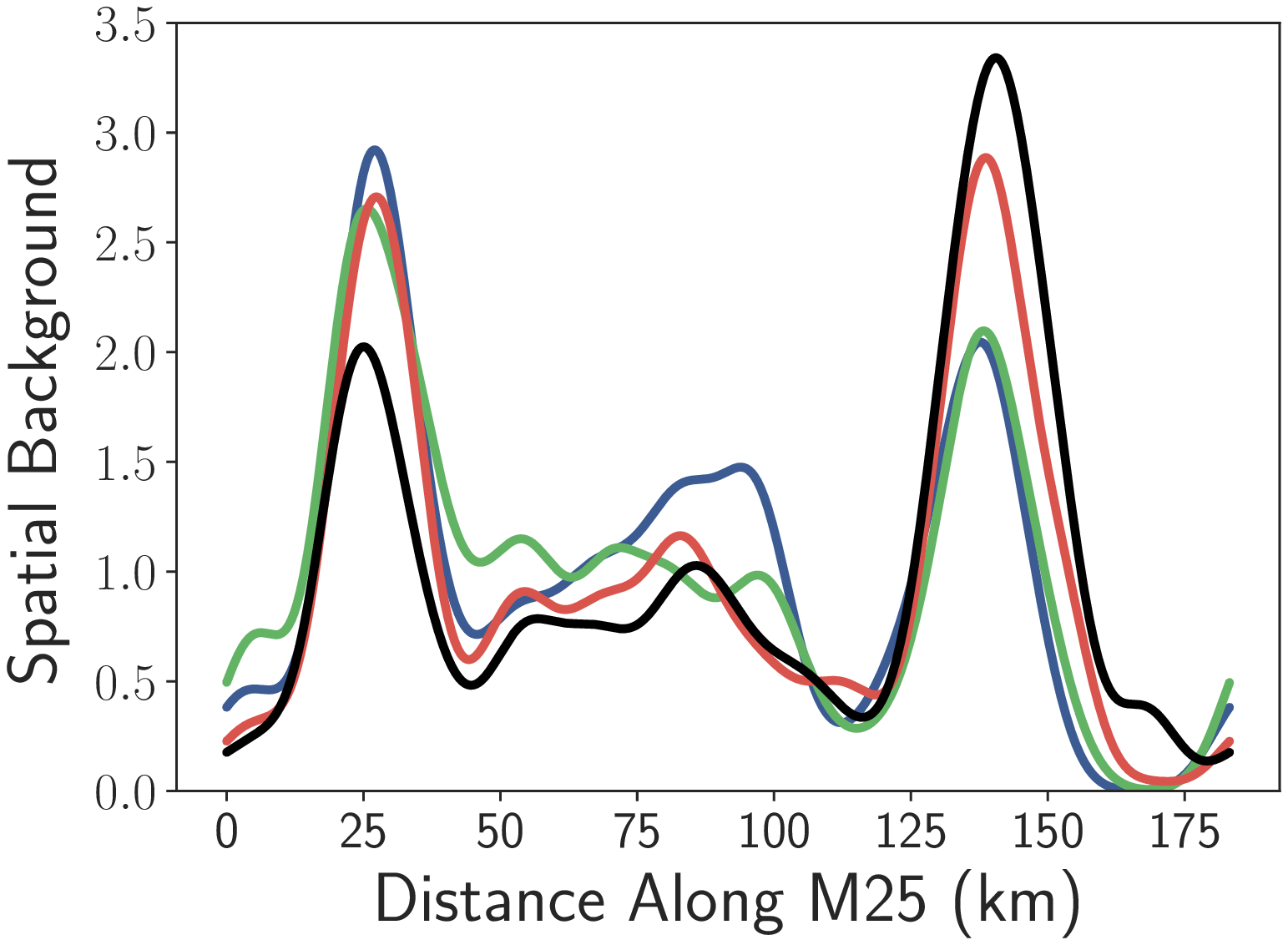}\label{fig:SpatialBackground3Months}}

     \subfloat[][Temporal Triggering]{\includegraphics[width=0.31\textwidth]{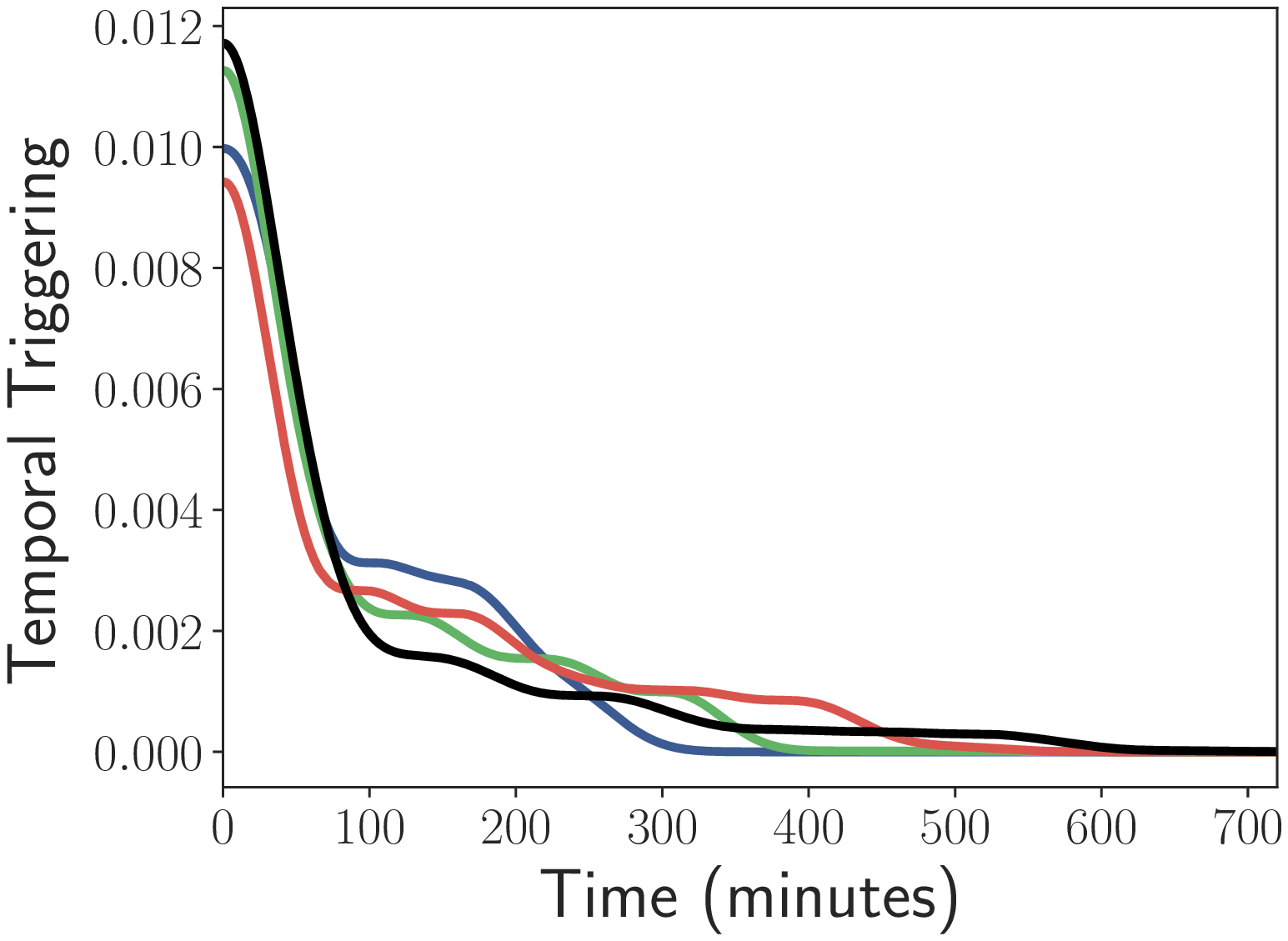}\label{fig:TemporalTrigger3Months}}
     \subfloat[][Spatial Triggering]{\includegraphics[width=0.31\textwidth]{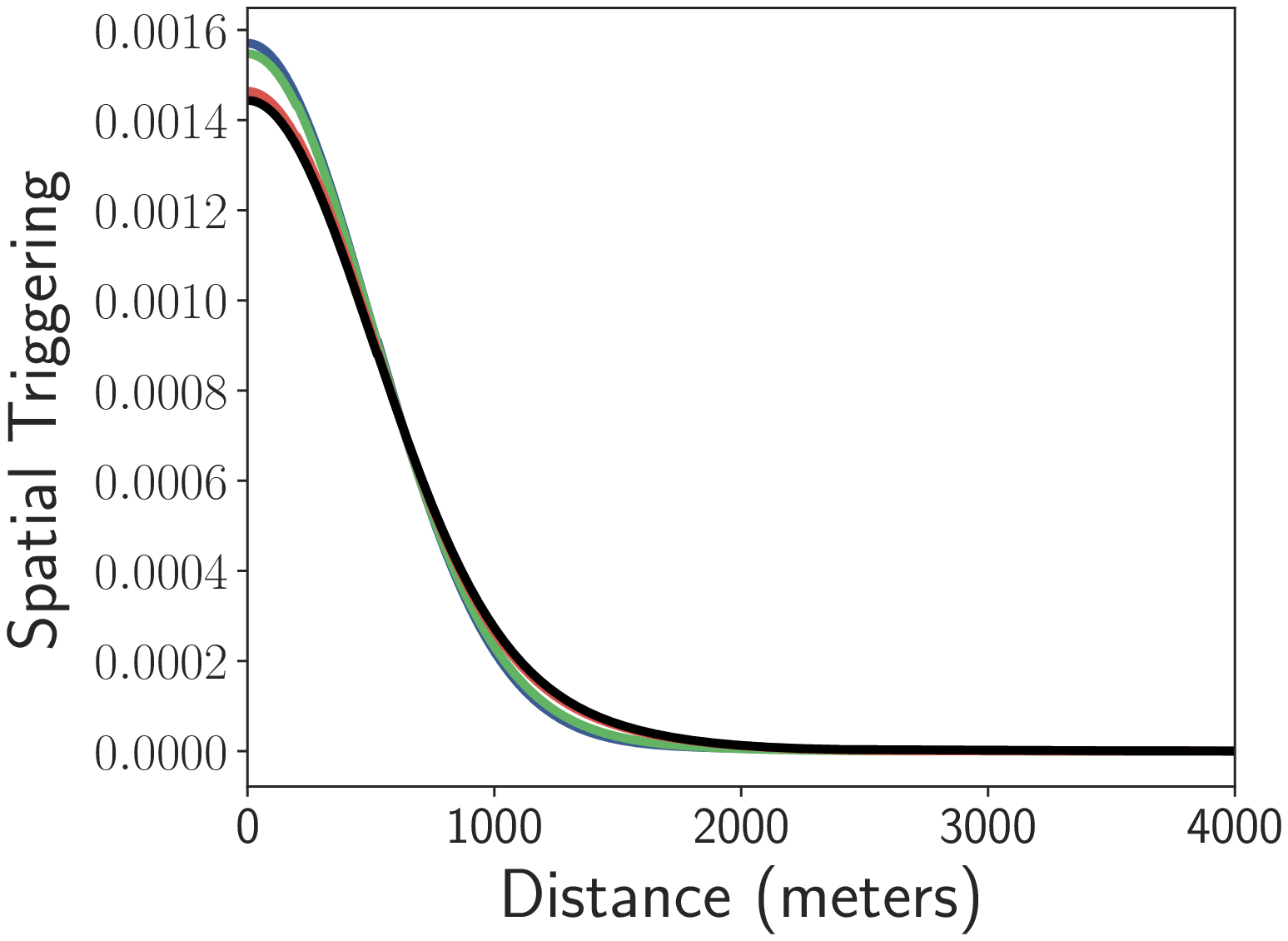}\label{fig:SpatialTrigger3Months}}

     \caption{Background and triggering components compared across different 3-month datasets. Datasets are: 09/2017-11/2017: {{\DenimBlueFull}}, 12/2017-02/2018: {{\BoringGreenFull}}, 03/2018-05/2018: {{\PaleRedFull}} and 06/2018-08/2018: {{\BlackFull}} (colour online).}\label{fig:3MonthSubsetComparision}
\end{figure}

Given our results for the year of data, we can question how resilient the background and triggering components are by inspecting subsets of the data. 
To test this, we partition our data into 3-month seasonal periods, and fit the model to each subset.
We then overlay the components in Fig. \ref{fig:3MonthSubsetComparision}.
It is clear from our results in Fig. \ref{fig:3MonthSubsetComparision} that there is varying amounts of consistency in components of the model.
Generally, we see in Fig. \ref{fig:DailyBackground3Months} that the daily background component is constant throughout the year, the only variation being in the 3-months of data starting on 03/2018, where the morning peak is a little more pronounced than any of the other datasets, and the evening is a little less.
It is difficult to make conclusions about the weekly component, when we have only 12 instances of each day of the week for a given 3-month subset.
Of particular interest is the spatial background through time, shown in Fig. \ref{fig:SpatialBackground3Months}, where we see that the peak around 140 kilometres along the M25 actually becomes more pronounced as time progresses.
Both the peak at 25 and 140 kilometres are present throughout all subsets of data, but later periods appear to show that the peak around Potters Bar is more significant later in the dataset compared to earlier. 
It is unclear if a physical change occurred leading to this, but would be of interest to investigate with more data.
Finally, the temporal and spatial triggering functions are generally quite consistent across all datasets.
We still see somewhat of a long decay in the temporal triggering, but time-scales of around 100 minutes remain.
Considering the triggering in each of these subsets, we attain $A$ values of 0.034, 0.063, 0.068 and 0.075 for the periods starting 2017/09, 2017/12, 2018/03 and 2018/06 respectively. 
Without more data, we cannot say if these values vary throughout the year, or if they are increasing over time.

\begin{figure}[!ht]
     \centering
     \subfloat[][Out of sample CDF Validation, fit to data 12/2017-02/2018, validated on data for 03/2018.]{\includegraphics[width=0.48\textwidth]{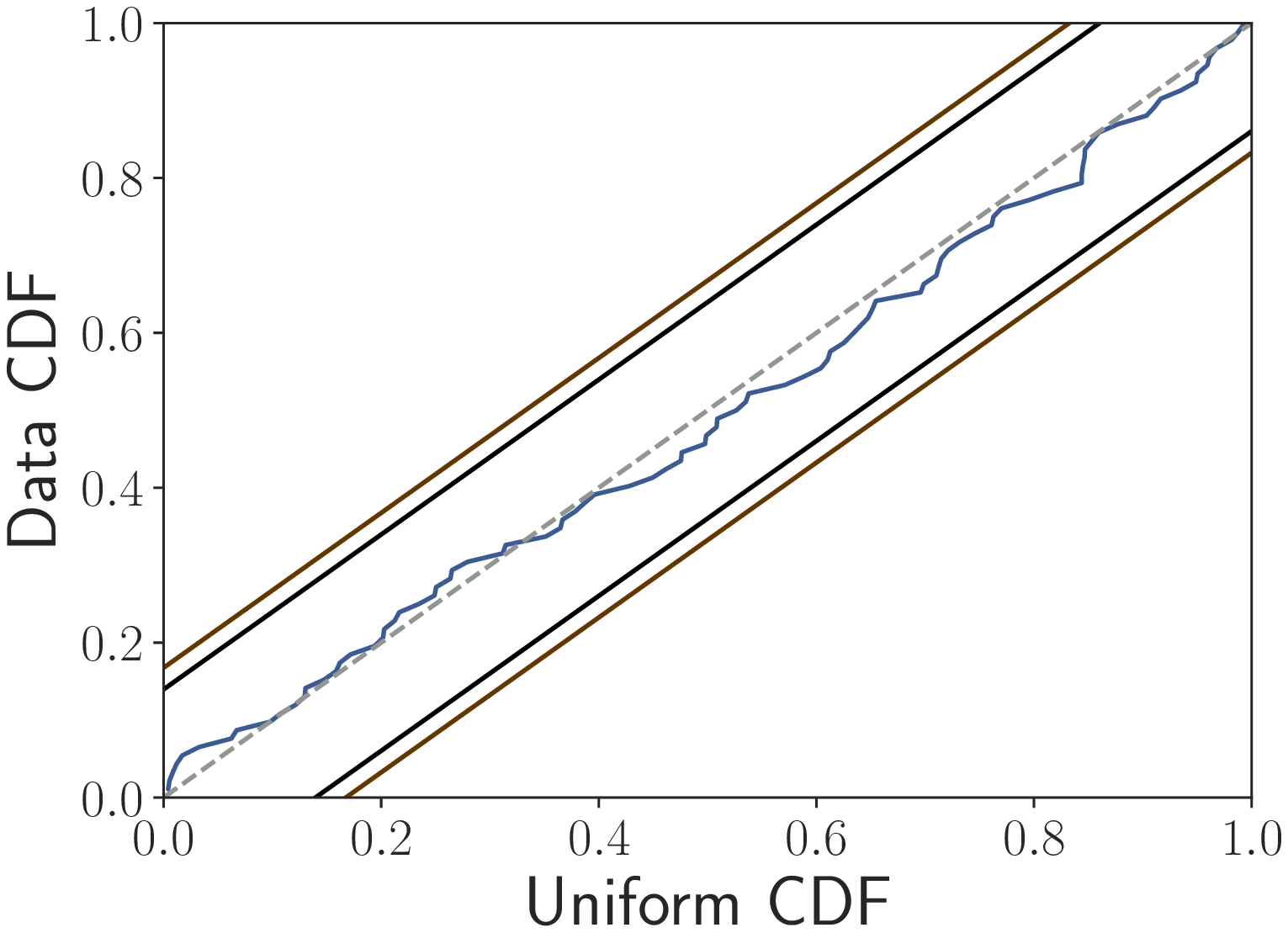}\label{fig:Dataset2CDF3MonthsOutOfSample}}
     \hfill
     \subfloat[][Out of sample CDF Validation, fit to data 06/2018-08/2018, validated on data for 09/2018.]{\includegraphics[width=0.48\textwidth]{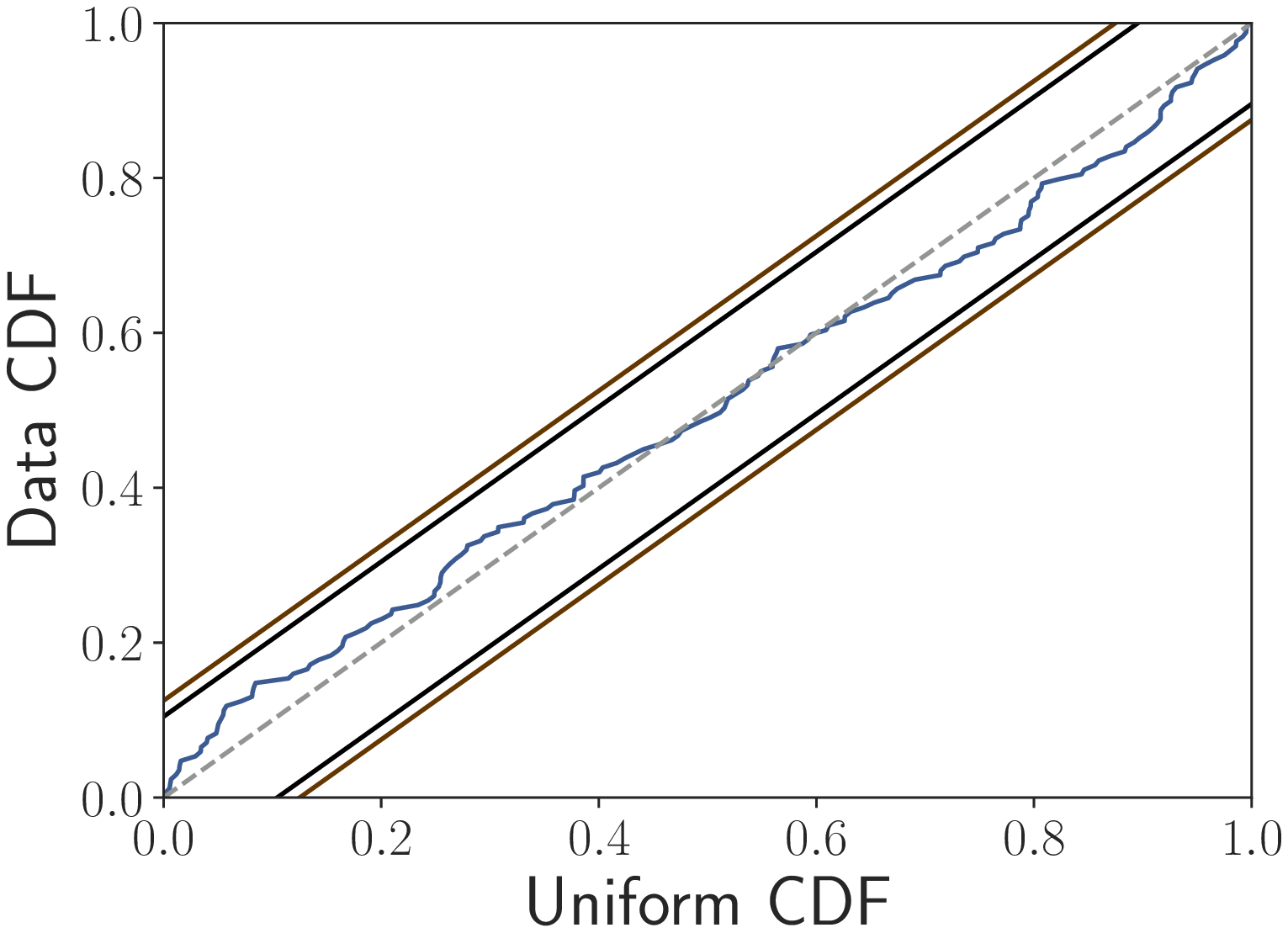}\label{fig:Dataset4CDF3MonthsOutOfSample}}
     \caption{Out of sample model validation for two datasets. Model results are shown with {{\DenimBlueFull}}, 95\% limits: {{\BlackFull}}, 99\% limits: {{\BrownFull}} and a reference line: {{\GreyDashed}} (colour online).}\label{fig:OutOfSampleValidation}
\end{figure}

We note that, for small temporal subsets of data, the trend component becomes less impactful in the model.
Recalling Fig. \ref{fig:SpatiaTrendBackground12Months}, the trend is almost flat for long periods, suggesting we can omit it for some temporal subsets.
As a result, we can perform out-of-sample model validation, something we are not aware of being done previously in the literature for this type of model.
Since the trend is omitted, and we see the triggering components are consistent through time, we can train the model on some 3-month subset of data, then perform our validation on unseen data.
We show two examples of this in Fig. \ref{fig:Dataset2CDF3MonthsOutOfSample} and \ref{fig:Dataset4CDF3MonthsOutOfSample}.
It is clear from Fig. \ref{fig:OutOfSampleValidation} that on short time-scales, one can use the model for acceptable performance out-of-sample.
This is not possible over longer time-scales however, due to the clear trend and varying spatial background evident in the data.

\subsection{Do Components Change for Significant Events?}

\begin{figure}[!ht]
     \centering
     \subfloat[][Daily Background]{\includegraphics[width=0.31\textwidth]{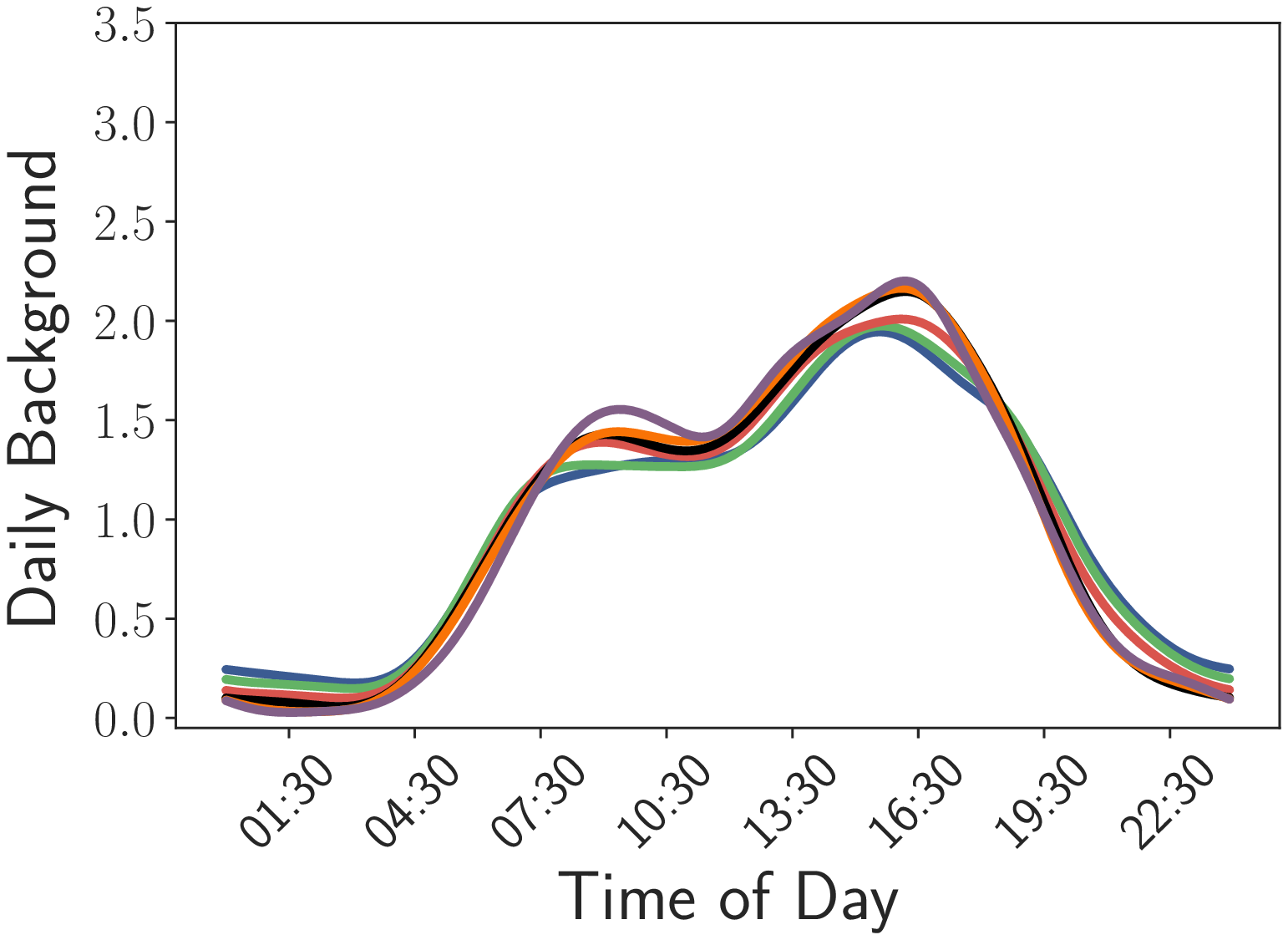}\label{fig:DailyBackgroundSigEvents}}
     \subfloat[][Weekly Background]{\includegraphics[width=0.31\textwidth]{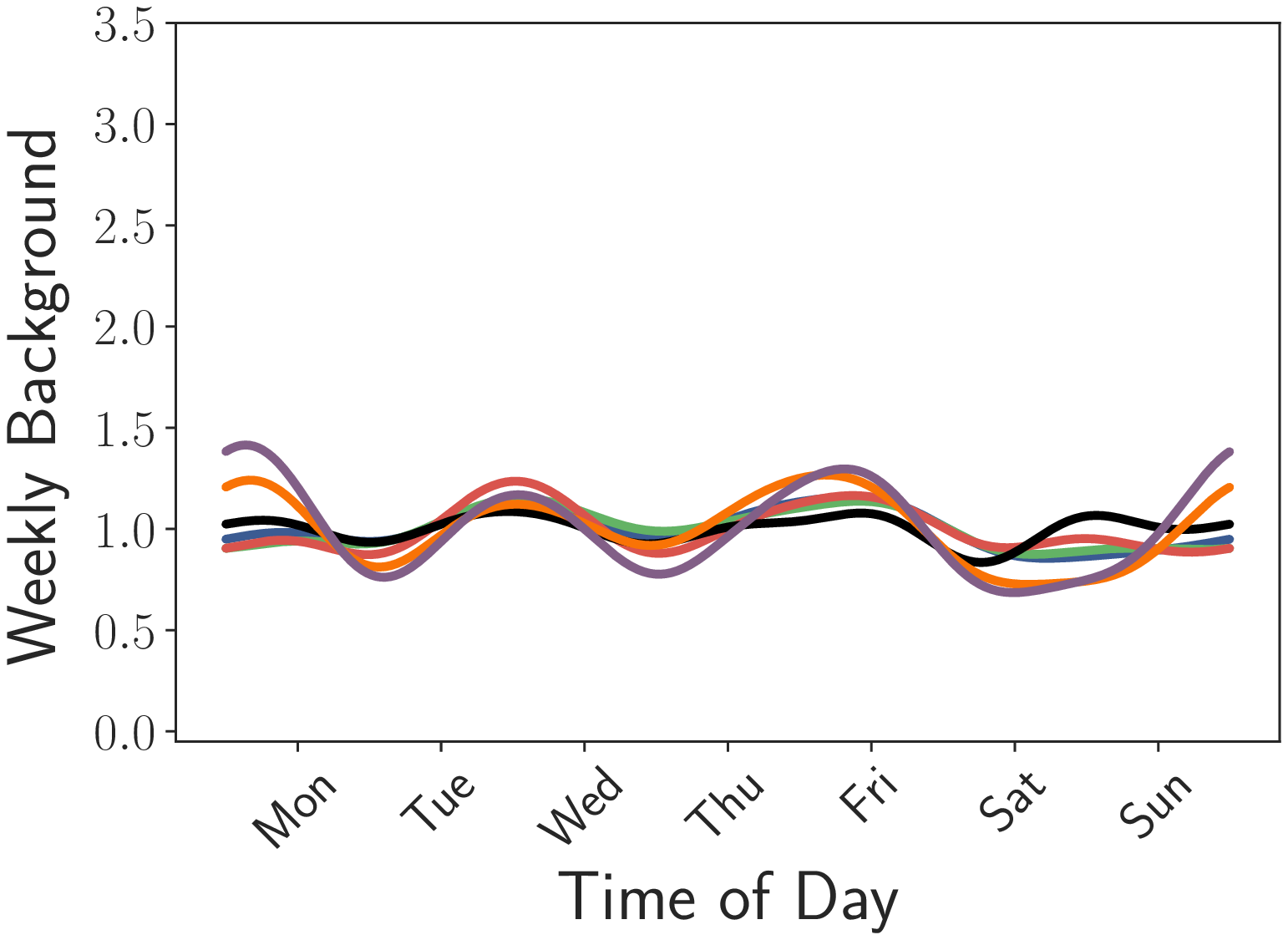}\label{fig:WeeklyBackgroundSigEvents}}
     \subfloat[][Spatial Background]{\includegraphics[width=0.31\textwidth]{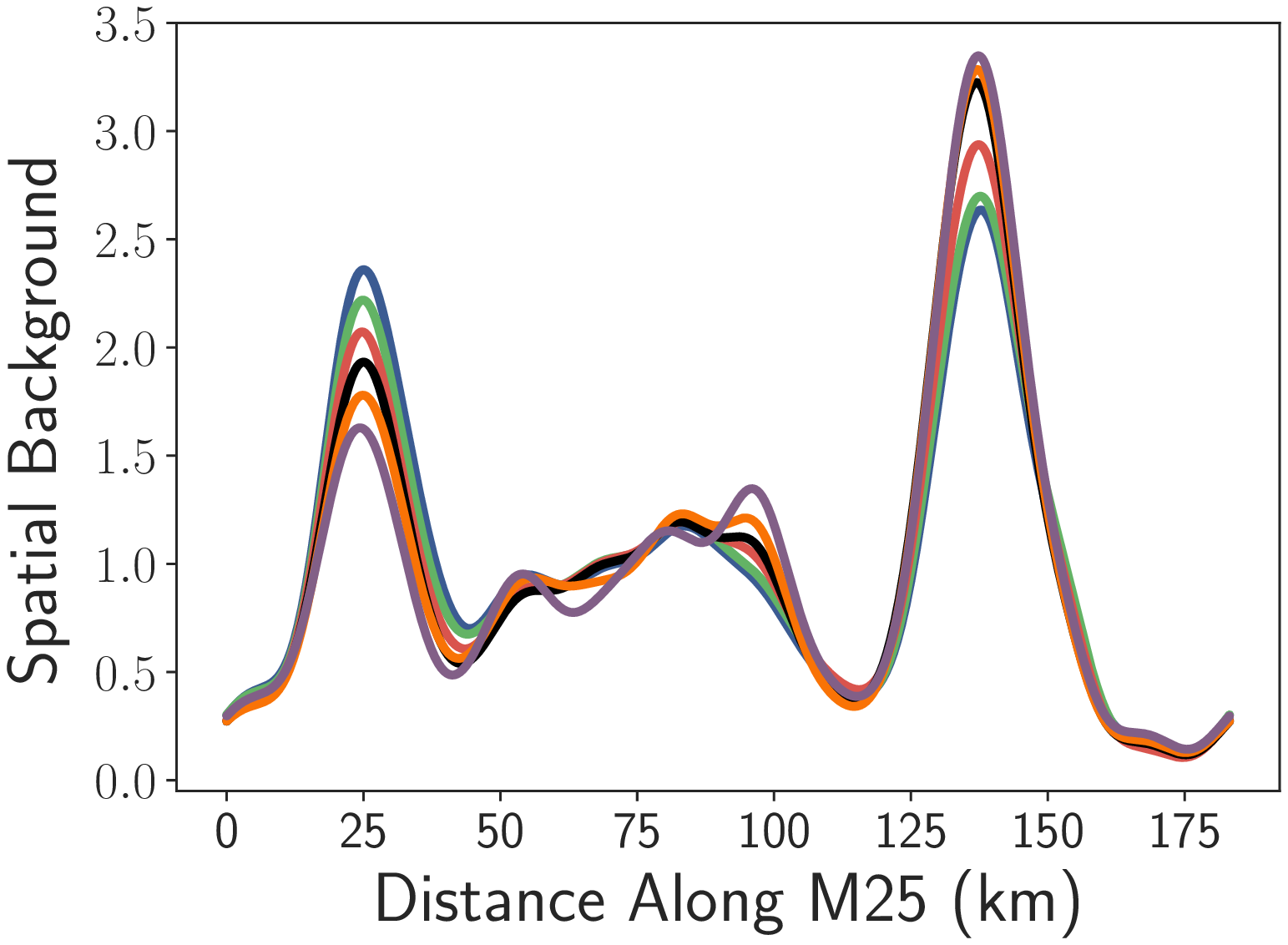}\label{fig:SpatialBackgroundSigEvents}}

     \caption{Background components compared across significant events, varying what speed decrease is required to define an event as significant. Thresholds are: 0\%: {{\DenimBlueFull}}, 10\%: {{\BoringGreenFull}}, 20\%: {{\PaleRedFull}}, 30\%: {{\BlackFull}}, 40\%: {{\OrangeFull}} and 50\%: {{\DustyPurpleFull}} (colour online).}\label{fig:ThresholdBackgrounds}
\end{figure}

Whilst all events flagged in NTIS should correspond to actual traffic incidents, many of them may not have had a significant impact on the traffic state. 
Consider the case where two vehicles have a minor collision, create no debris, and the drivers pull into the closed hard-shoulder to exchange insurance information. 
In such a case, there should be little impact on flow, travel time and speed.
Additionally, if a vehicle breaks down and pulls into the hard-shoulder, and the road is far from capacity, we should again see little drop in average speed.
To consider only the behaviour of events that have some significant impact, we inspect the link-level data, which contains significantly less noise than the loop-level.
For a given event window, we consider the largest percentage drop in speed between a simple historical median segmentation profile and measured values across the entire window.
If this percentage drop is above some threshold, then we say the event caused a significant impact on the traffic state.
Further discussion of this seasonal model is given in the supplementary material.
As we raise this threshold, we isolate more extreme events but discard so much data that it is no longer reasonable to fit a model.
To retain enough data for fitting, we consider only thresholds between 0 and 50\%.
We split our dataset into subsets containing only events that lead to a speed decrease of at-least 0\%, 10\%, $\dots$, 50\%, then re-run our model fitting on these subsets.
The resulting background components are visualized in Fig. \ref{fig:ThresholdBackgrounds}.

Fig. \ref{fig:DailyBackgroundSigEvents} shows the daily background is reasonably stable across different thresholds of significance. 
As we raise the threshold, the morning and evening peak structure becomes clearer, and the periods very early and late in the day are lowered.
Fig. \ref{fig:WeeklyBackgroundSigEvents} shows that weekends have lower intensity than weekdays as we raise the threshold for significance.
This is likely due to demand being significantly lower on weekends compared to weekdays, and hence when an event does occur, there is less chance of an queue forming as the road is further from capacity than on a weekday.
Finally, the spatial background in Fig. \ref{fig:SpatialBackgroundSigEvents} clearly shows that the second peak, identified around 140 kilometres along the M25, appears to experience more significantly impactful events than the first peak, observed around 25 kilometres along the M25.
This suggests that not only is Potters Bar a `hot-spot' for events, but also that the events here are some of the more extreme on the network.
Analysis of triggering during different significance thresholds shows that the time-scales and $A$ values remain consistent throughout. 
The only visible difference in the triggering functions is that for larger thresholds, the functions decay more quickly.  

\section{Conclusion}\label{sec:Conclusion}

We have analysed the spatio-temporal variation in the incident rate on London's M25 motorway over a period of one year using a model that distinguishes between primary and secondary events.
This variation is found to be strongly inhomogeneous.
The temporal variation shows a strong daily double peak structure reflecting commuting patterns superimposed in a weaker weekly variation with a peak on Fridays and a trough on Saturdays. 
This pattern of temporal variation remains stable over the data period.
The spatial variation shows two primary peaks in intensity. 
The first and largest is in the vicinity of the Potters Bar Interchange.  
The other is in the vicinity of Junctions 5 and 6  where the M26 and M23 join the M25. 
The peak at Potters Bar appears to increase in intensity during the data period, and is more pronounced when we condition on the most  significant events in terms of impact on traffic speed.
We find that 6-7\% of the observed incidents are most probably secondary incidents under the assumptions of our model.
Plausible time and length scales emerge for the range of the triggering effects: 100 minutes in the temporal triggering, and 1 kilometre for spatial triggering. 
From these figures we conclude that the effects of secondary incidents is a small but detectable feature of the M25 incident data set.
We suggest that, on the M25, the scope to further reduce incident rates by reducing secondary incidents is limited compared to what could be achieved by reducing the peaks at specific times or `hot-spot' locations.

We have extended existing work to limit the freedom of the self-excitation components for non-parametric Hawkes processes, along with considering unidirectional spatial triggering.
We further proposed an out-of-sample validation method which is viable when the trend component is roughly constant over time.
Applying this to our dataset showed we have avoided over-fitting the model, despite its significant freedom.
These advancements are not specific to the application discussed here, and can be used when models of this form are applied to different domains with similar practical considerations.
The modelling framework is novel in the context of traffic incident analysis, and our methodological refinements ensure the model captures relevant properties of the application.

In terms of further work, it would be informative to repeat our analysis for a major road without the MIDAS system or smart motorway features since we expect this infrastructure to reduce the risk of secondary incidents.
Additionally, it would be interesting to apply this analysis to urban road network, however directly extending our methodology to account for this structure appears difficult.
Whilst smoothing of point-processes on a network has been explored recently, for example in \cite{first_and_second_order_characteristics_of_spatio_temporal_point_processes_on_linear_networks}, further complication remains regarding smoothing of the triggering functions on a network.
When we take data-points in $\mathbb{R}^N$ and compute distances between them, we attain a value in $\mathbb{R}^+$, and differences between these distances lie in $\mathbb{R}$.
However, the same is not necessarily true for differences between distances computed on a network, which still attain real values, but belong to the sub-space stemming from their network-constrained spatial structure. Taking this into account makes smoothing the triggering functions complex and deserves further attention in future work.
General comments about clustering of events on urban roads, made in \cite{network_based_likelihood_modeling_of_event_occurances_in_space_and_time_a_case_study_of_traffic_accidents_in_dallas_texas}, suggest that secondary incidents remain a feature of urban roads.
One could also extend our out-of-sample validation to include a trend component.
To do so, a parametric form of the trend is required, as we cannot extrapolate our current non-parametric estimate.
Finally, one could incorporate more variables that may influence traffic incidents, for example weather conditions.

\section*{Acknowledgements}
We thank Dr. Steve Hilditch, Thales UK for sharing expertise on NTIS and UK transportation systems. This work was supported by the EPSRC (grant number EP/L015374/1).

\appendix

\section*{Supplementary Material}

\section{Problems With Link Specification of Events}\label{appendix:ProblemsWithLinkSpecification}

To understand why we can localize events, we detail the two scales of data being provided by NTIS.
The first is at the loop-level, with time-series of speed, flow and occupancy being reported each minute for each sensor in the network.
The second is at the link-level, where aggregated data from all sensors that lie on a link is provided each minute, specifically speed, flow, occupancy and travel time.
We extract both of these sets of data across our studied domain, and apply a 5-minute rolling average to smooth out the time-series. 

\begin{figure}[!ht]
     \centering
     \subfloat[][Distribution of link lengths through the M25.]{\includegraphics[width=0.48\textwidth]{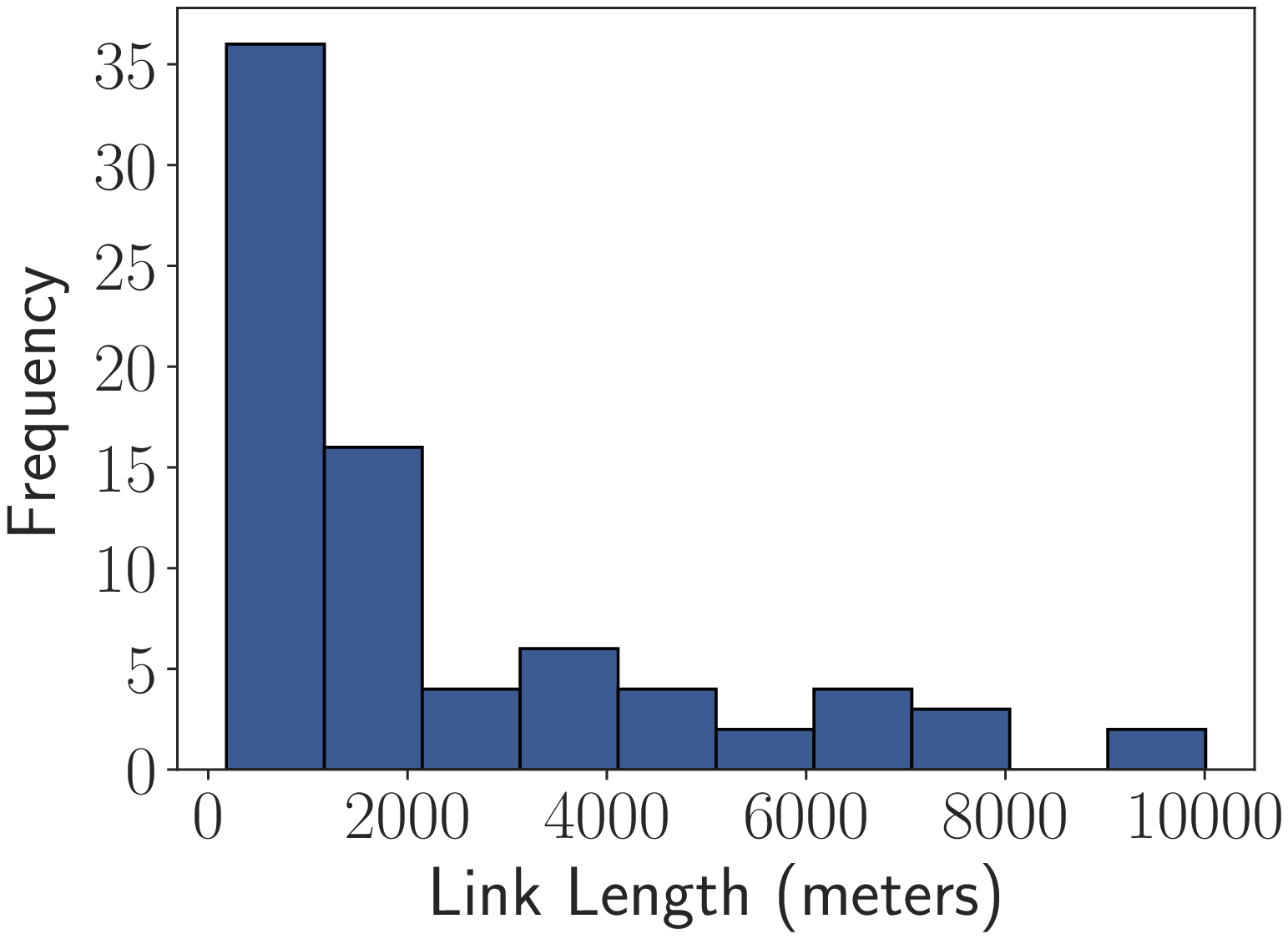}\label{fig:LinkLengthDist}}
     \hfill
     \subfloat[][Distribution of gaps between loop sensors on the M25.]{\includegraphics[width=0.48\textwidth]{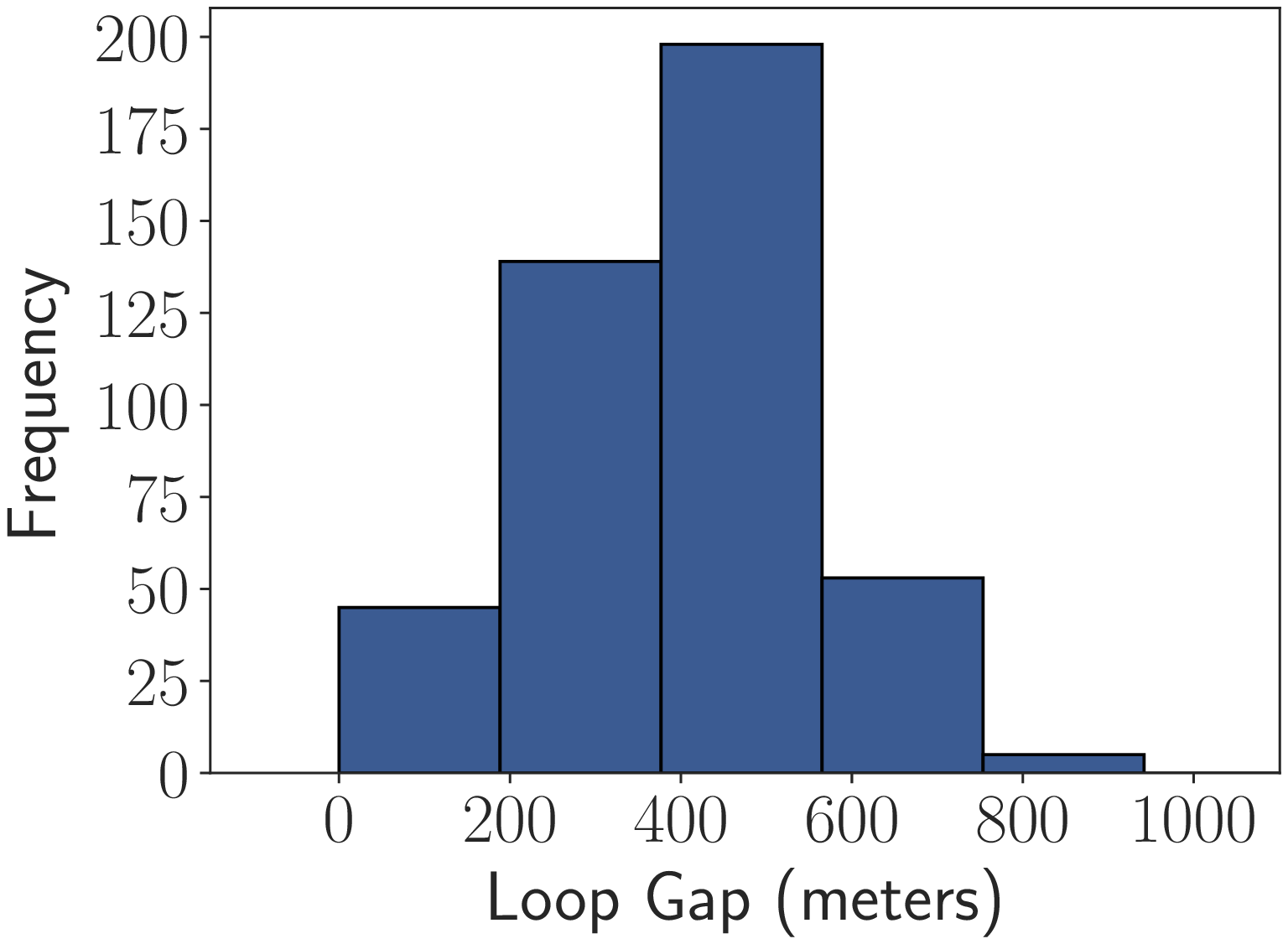}\label{fig:LoopGapDist}}
     \caption{Comparison of link lengths and distance between successive loop sensors. We see that there is wide range of link lengths, with a small number being close to 10 kilometers in length.  On the other hand, distances between successive loop sensors are highly concentrated between 0 and 800 meters.}\label{fig:LinkAndLoopSizeDist}
\end{figure}

When an event is recorded on our network, we are told which link it has occurred on, however NTIS links vary vastly in size, with the distribution of link lengths given in Fig. \ref{fig:LinkLengthDist}, along with the distribution of gaps between successive loop sensors across the entire network in Fig. \ref{fig:LoopGapDist}.

The wide range of link lengths shown in Fig. \ref{fig:LinkLengthDist} clearly indicate that further localization needs to be performed to narrow down a specific point on the link before we can apply point-process methodology.
However, there is hope that the loop sensors are fine-grained enough in space to provide an estimate of this, with Fig. \ref{fig:LoopGapDist} showing that any point in the network should have a loop sensor less than 400 meters away.
This is a far higher level of detail than initially provided by NTIS events, so we decide to refine our data further by using the time-series provided by the loop sensors.
This is not a trivial task however as we do not have labeled data on which we can train a model, and instead must define sensible criteria that indicate the location of an incident. 
This is an extensive task, and detailed in section \ref{sec:SecLocalizingEvents}

\subsection{Event Localization Methodology}\label{sec:SecLocalizingEvents}

We note from the outset that detection of traffic events in both space and time is an active research field, and various approaches are being considered using an increasing variety of data sources.
We are in an atypical situation in that we know a temporal and spatial window in which an event occurred, and are only searching for a more fine-grained spatial location within this window.
As such, we do not aim to completely solve the event detection problem using inductive loop data, rather we try and develop an effective methodology to take a given window with a known event in and argue what single set of sensors the event may lie between.
From discussions with industry experts, we consider the following properties to be clear signatures of a significant traffic event:
\begin{itemize}
\item[a)] upstream of an event location, speed will be decreased and occupancy will be increased compared to the seasonal values
\item[b)] downstream of an event location, speed may still be decreased, but less so than upstream of the accident
\item[c)] downstream of an event location, occupancy will be decreased compared to the seasonal values
\end{itemize}

Given the industry expert criteria, we first develop simple seasonal models for each loop sensor in our network. 
There is clear seasonality on the weekly scale in traffic data, with commuting days in the U.K. being Monday to Friday, and Saturday and Sunday having less vehicles on the road.
Taking this as the leading seasonal component, we construct a simple seasonal model by taking all data on a given weekday at a given time of day, and then finding the median value, using this as a reasonable seasonal estimate.
Doing so, we have a model for each weekday, at the minute level, fit to each loops data separately.
We produce one such seasonal model for each traffic variable on a loop.

\begin{figure}[!ht]
    \centering
    \subfloat[][Simple Diagram of the link with sensors marked ({{\DenimBlueDashed}}), the localized event location ({{\BoringGreenDotted}}), and the start and end of the link ({{\PaleRedFull}}).]{\includegraphics[width=0.64\textwidth]{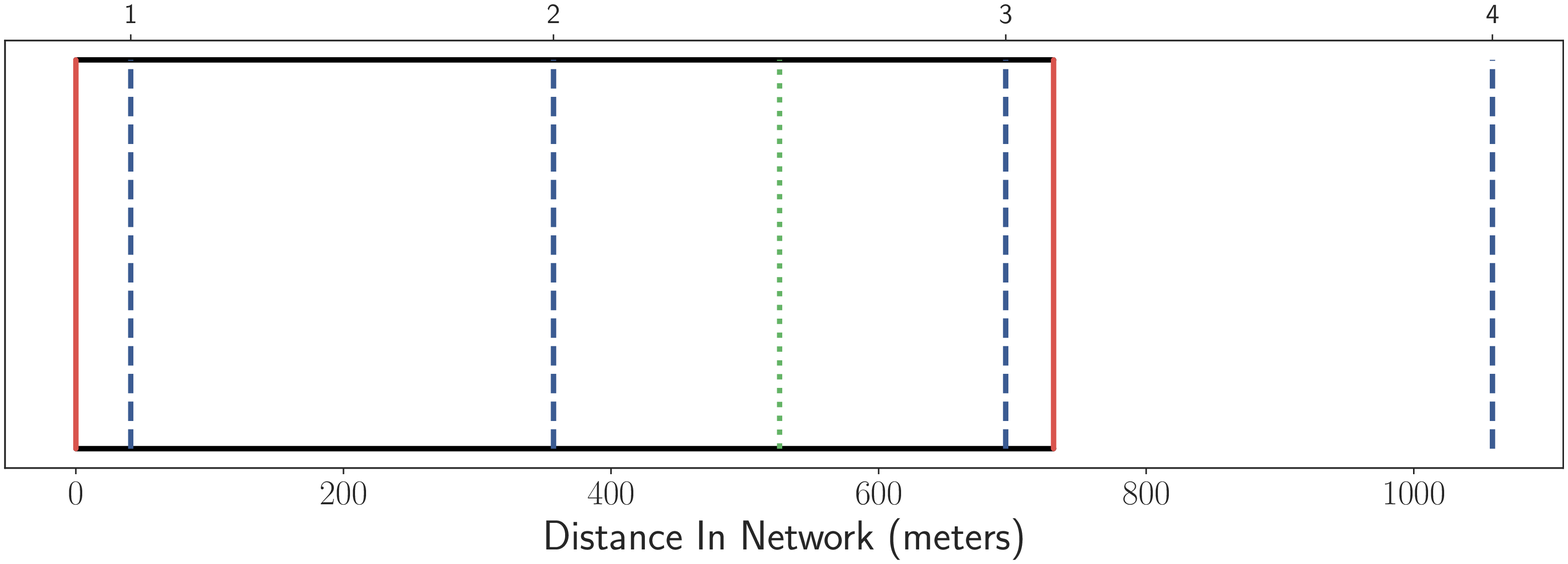}\label{fig:LocalizationExample1LinkPlot}}

    \subfloat[][Loop 1 - speed]{\includegraphics[width=0.31\textwidth]{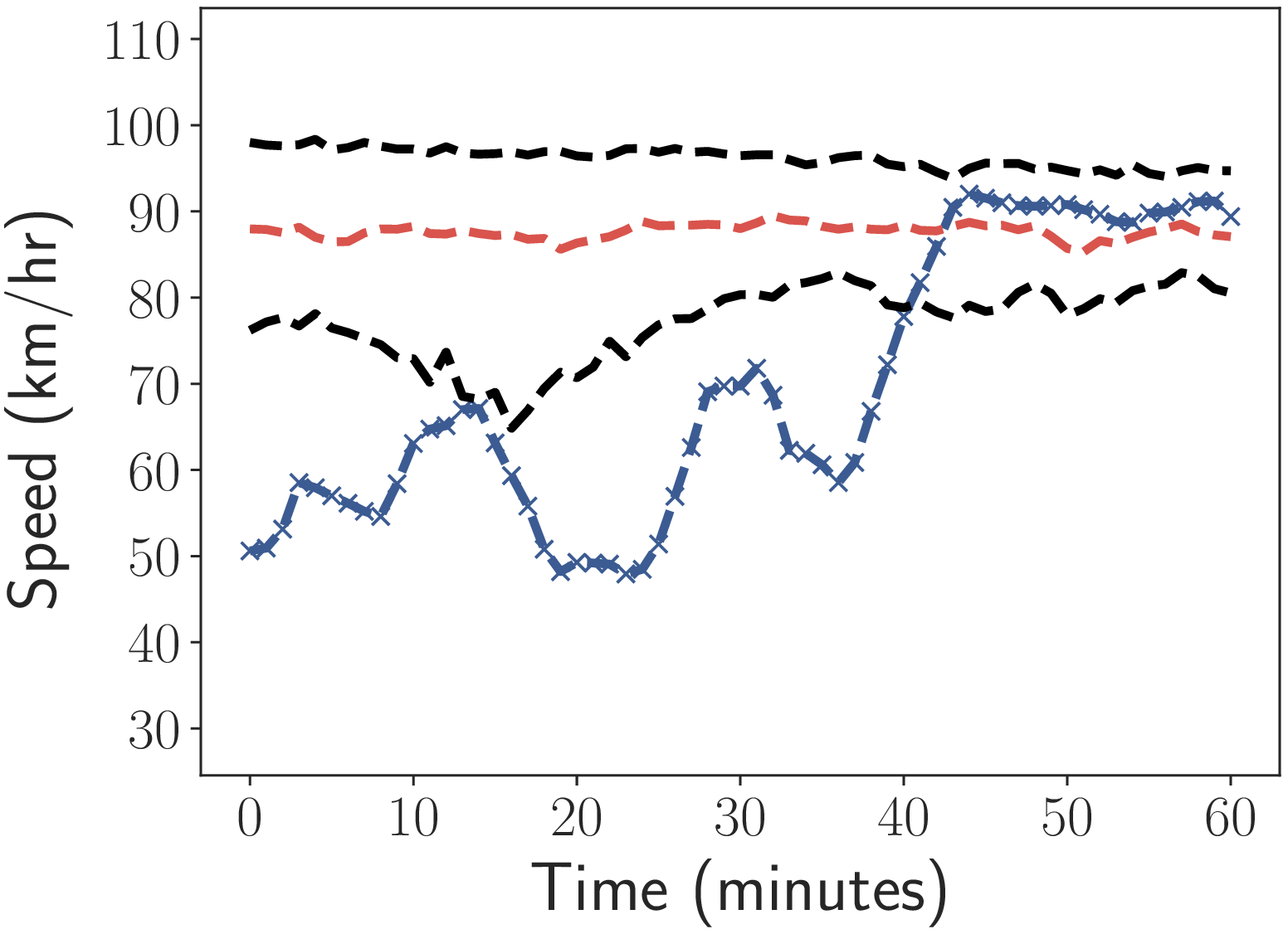}\label{fig:LocalizationExample1Speed1}}
    \subfloat[][Loop 2 - speed]{\includegraphics[width=0.31\textwidth]{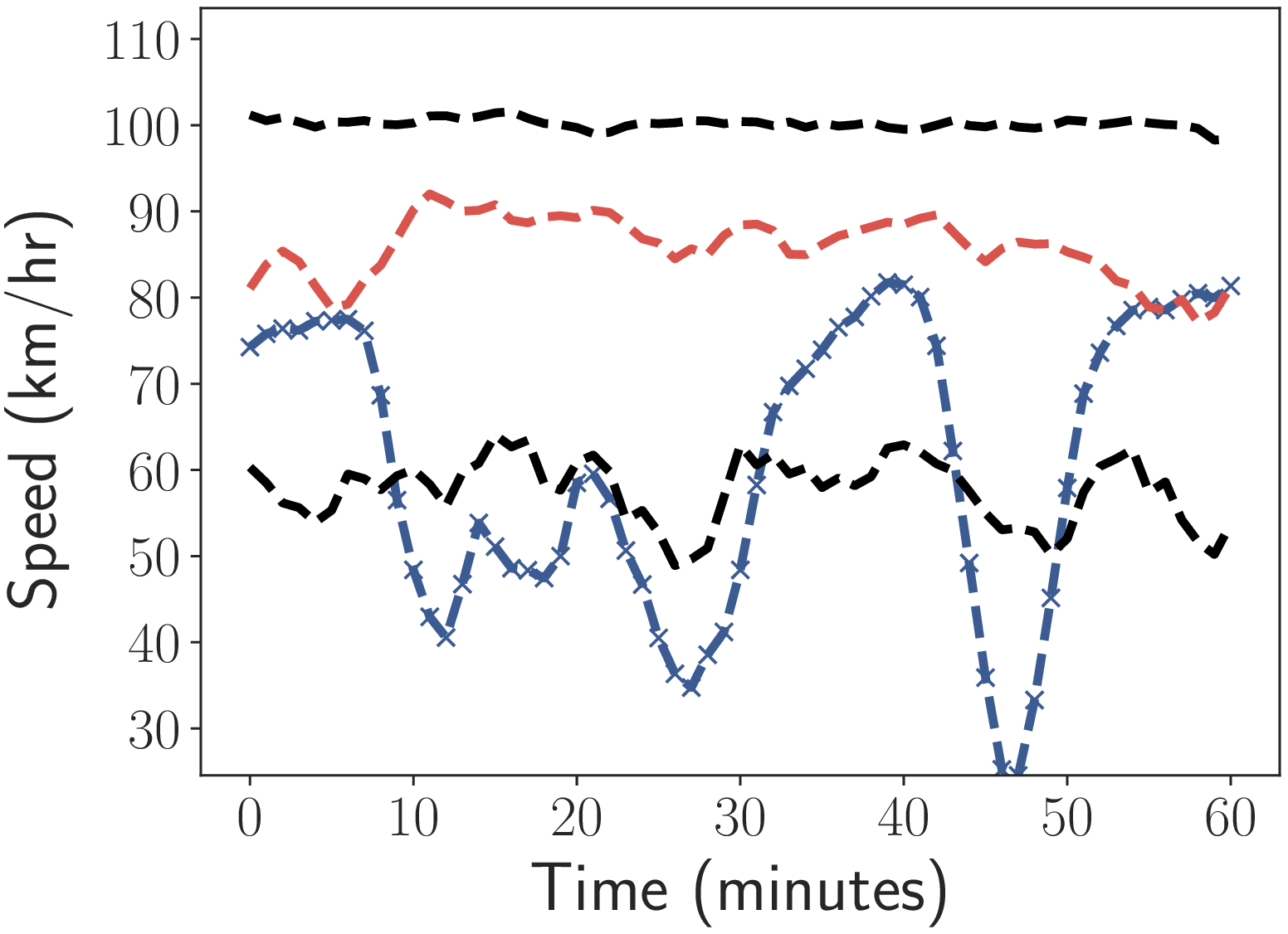}\label{fig:LocalizationExample1Speed2}}
    \subfloat[][Loop 3 - speed]{\includegraphics[width=0.31\textwidth]{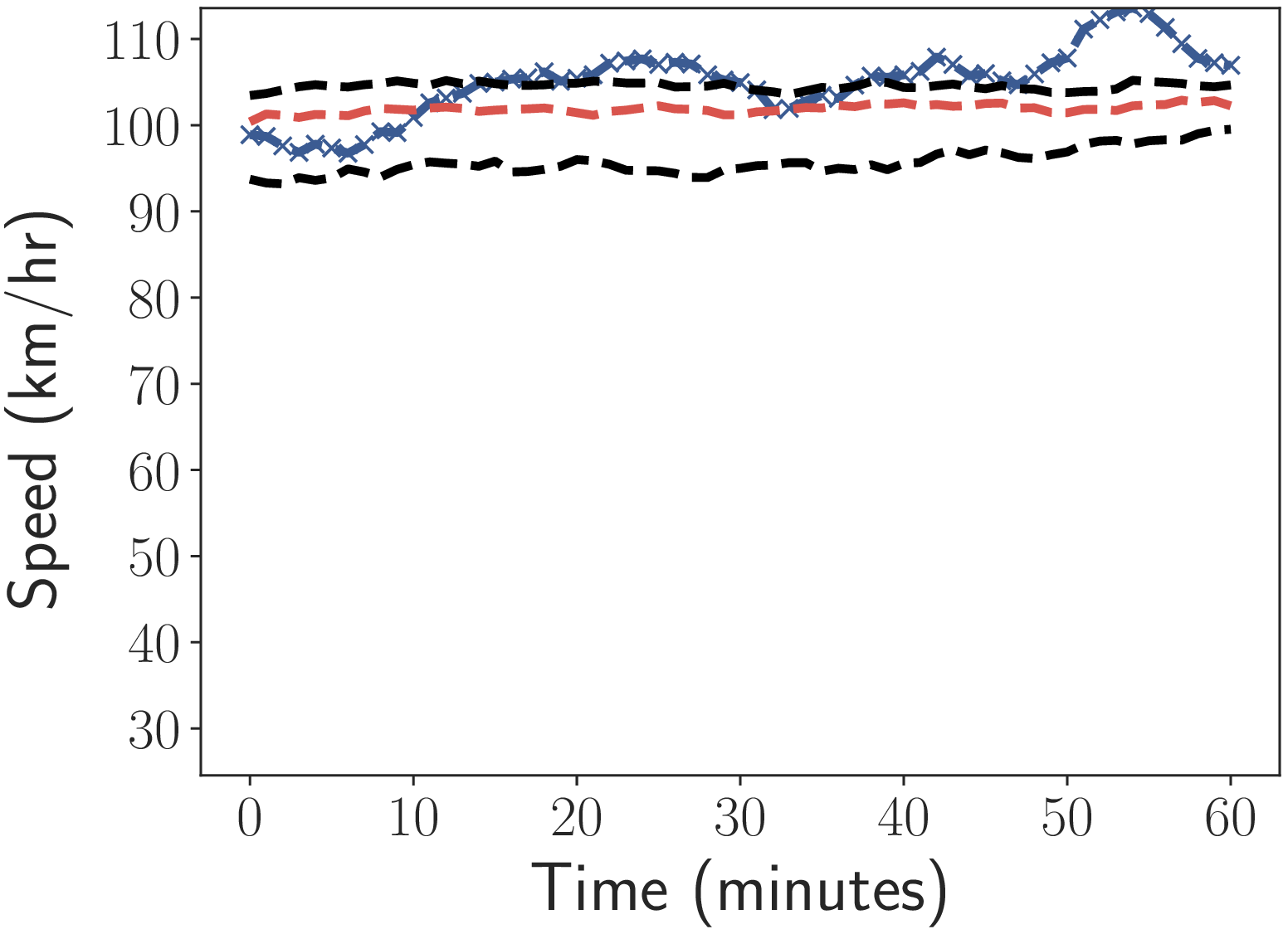}\label{fig:LocalizationExample1Speed3}}

    \subfloat[][Loop 1 - occupancy]{\includegraphics[width=0.31\textwidth]{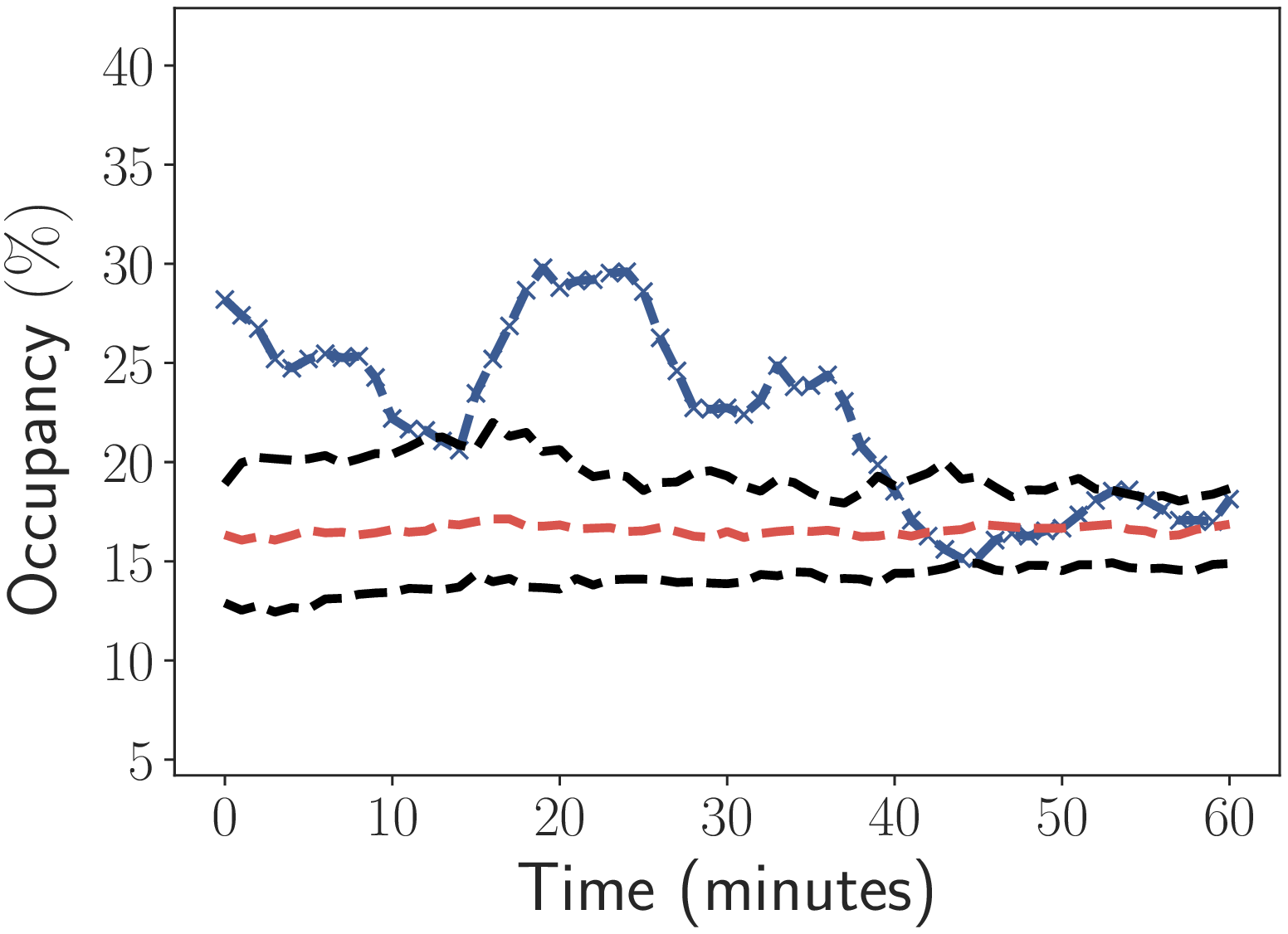}\label{fig:LocalizationExample1Occupancy1}}
    \subfloat[][Loop 2 - occupancy]{\includegraphics[width=0.31\textwidth]{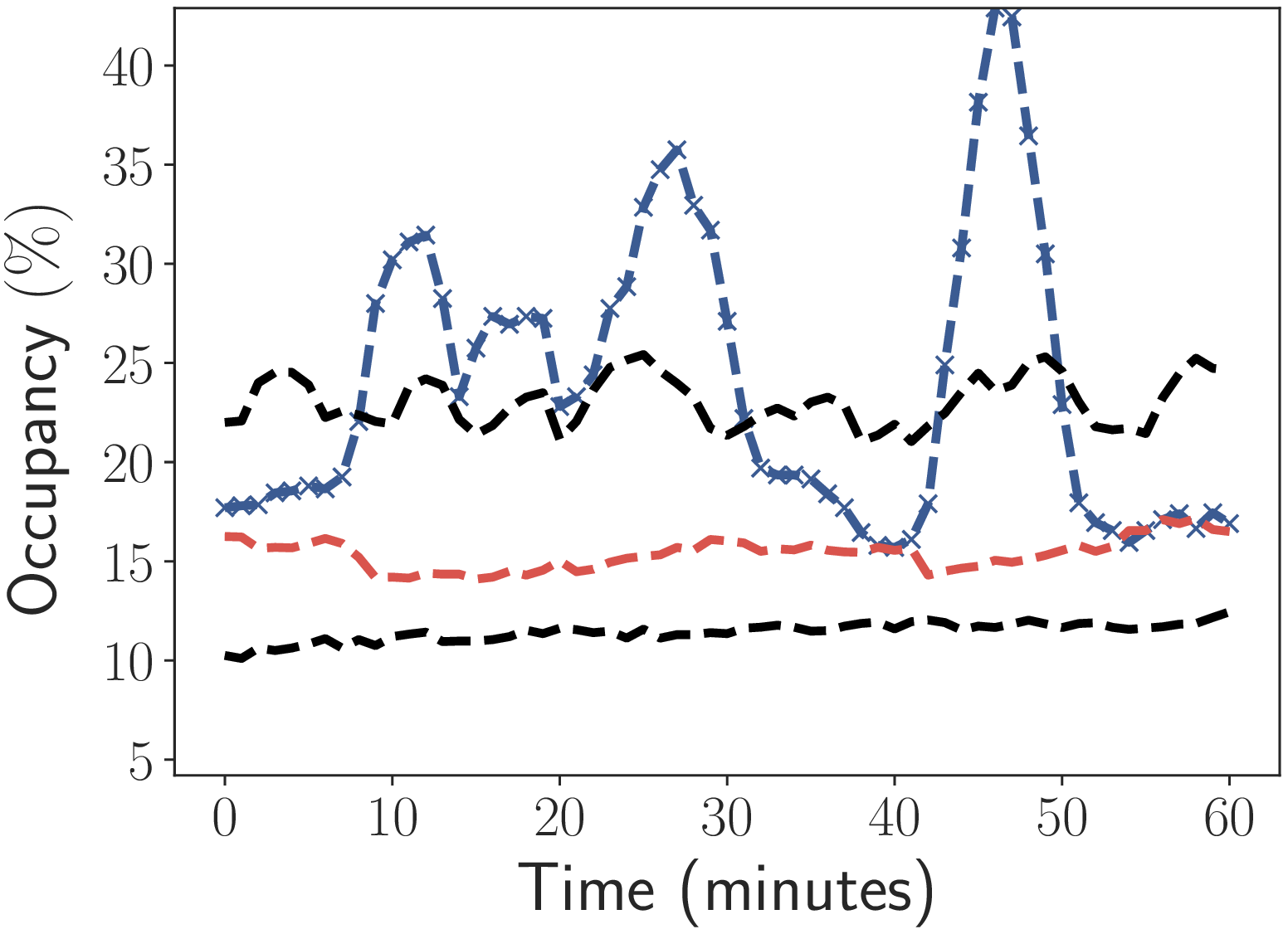}\label{fig:LocalizationExample1Occupancy2}}
    \subfloat[][Loop 3 - occupancy]{\includegraphics[width=0.31\textwidth]{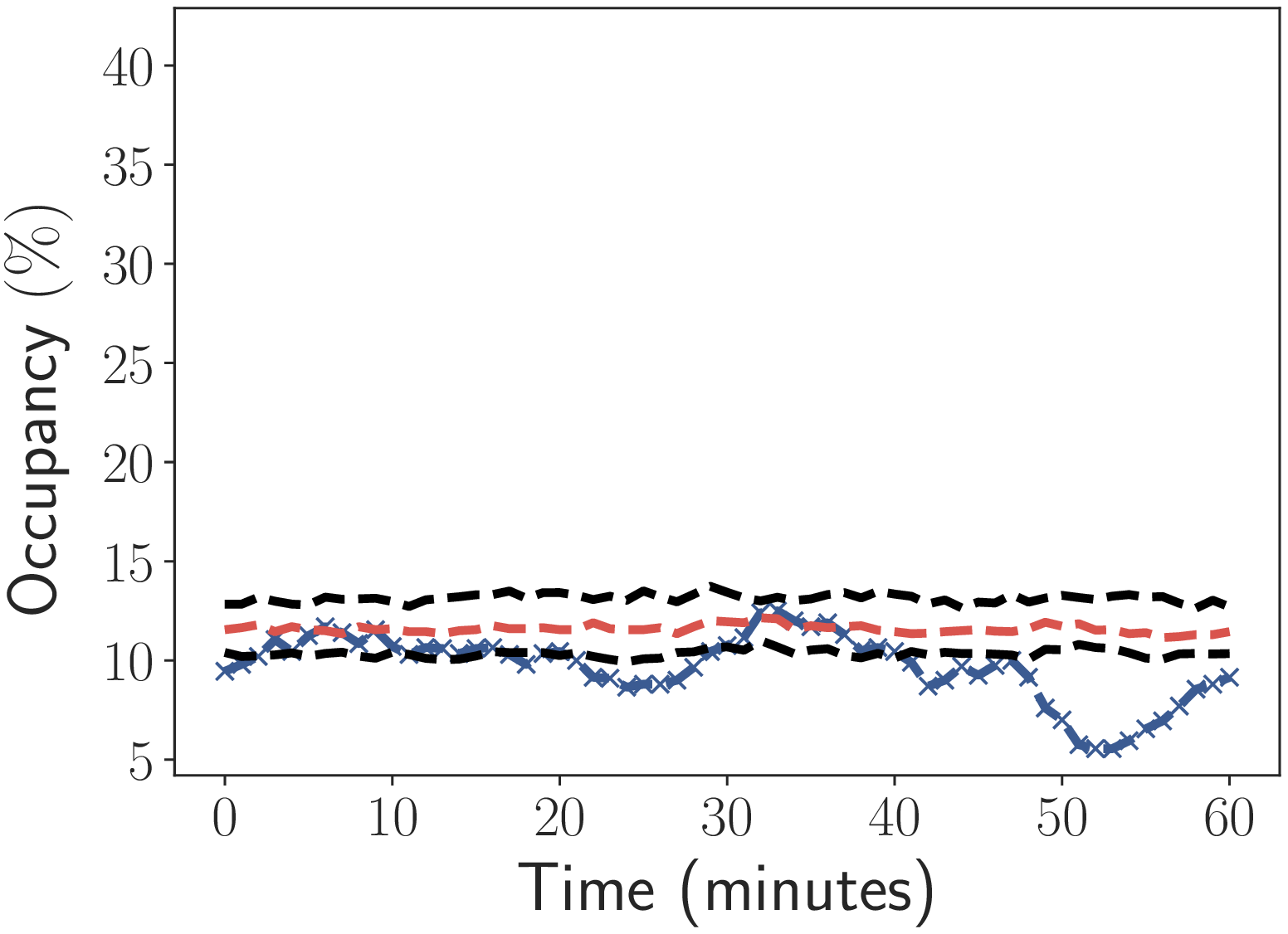}\label{fig:LocalizationExample1Occupancy3}}

    \caption{An example result of our localization procedure.  We plot the data ({{\DenimBlueDashed}}), seasonal median ({{\PaleRedDashed}}) and the 20th and 80th percentiles for the particular traffic variable ({{\BlackDashed}}). These give a sense of how much variation there is in the data. This is the first link in our network, so we show the next loop sensor along for a sense of scale. We see that sensors 1 and 2 have large drops in speed and increases in occupancy,  however sensor 3 appears reasonably seasonal. Our methodology has then placed the event between sensors 2 and 3. }\label{fig:LocalizationExample1}
\end{figure}

We then consider what is reasonable to develop with our available data. 
Ideally, we would design and validate some localization methodology incorporating spatial-temporal information from the loop data, inferring a location from the behaviour of all loops.
However, we have no data with the ground truth locations, so we cannot reasonably develop such a model.
Instead, we can use a simple `rule of thumb' approach based on existing methodology.
Numerous historic methodologies in traffic theory \cite{california_algorithm_original}, \cite{mcmaster_distinguising_between_incident_congestion_and_recurrent_congestion_a_proposed_logic} compare adjacent sensors to determine two points, one where the data appears to show an incident, and another where the data does not.
We can do the same, and given we know there must be an event in the given window, we can simply ask at what point do we see the largest discrepancy for two adjacent sensors.
First, we consider any two sensors in our network, $i-1$ and $i$ $\, \text{for all} \, \, i  \, \in \, \{ 2, \cdots, N \}$.
We denote the residual series for speed and occupancy on a sensor $i$ as $RS_i$ and $RO_i$ respectively.
We then compute the spatially differenced values:
\begin{equation}\label{equ:ResidualsPart1}
\Delta RS_i = RS_i - RS_{i-1}, \, \, \, \, \, \Delta RO_i = RO_i - RO_{i-1}.
\end{equation}
We then define an `event impact score' between the two loops as:
\begin{equation}\label{equ:EventImpactScore}
EIS_i = \Delta RS_i - \Delta RO_i.
\end{equation}
Since $\Delta RS_i$ will likely be positive and $\Delta RO_i$ will likely be negative if an event occurs between $i-1$ and $i$, then we look for the pair of loops at which the value of $EIS_i$ is largest, and place our localized event halfway between these two loops.
There is of course huge scope to improve upon such a model, but without data to validate more advanced approaches it is sufficient to provide a method that agrees with existing literature and common sense rules.
In Fig. \ref{fig:LocalizationExample1}  we plot an example of localizing an event based on our simple methodology applied to loop sensor data.

\section{Model Derivation}\label{appendix:Methodoloy}

Before detailing how one derives the model, we recall from the main text that the conditional intensity function takes the form: 
\begin{equation}\label{equ:BasicModelEquationSupp}
\lambda(t,x) = \mu_0\mu_d(t)\mu_w(t)\mu_t(t)\mu_s(x) + A\sum_{\substack{t_i < t \\ x_i > x}}g(t-t_i)h(x-x_i),
\end{equation}
and one can enforce monotonicity be solving: 
\begin{equation}\label{equ:OptimProblemMonotonicSupp}
\begin{aligned}
\min_{p_1, \dots, p_N} \quad & D_0(p_1, \dots, p_N)  \\
\textrm{s.t.} \quad & \frac{d\hat{\nu}_{mono}(x | p_1, \dots, p_N )}{dx} \leq \epsilon \\
					& p_i \geq 0 \, \, \, \text{for all} \, \, i \in [1, 2, \dots, N] \\
					& p_i \leq 1 \, \, \, \text{for all} \, \, i \in [1, 2, \dots, N] \\
					& \sum_{i=1}^N p_i = 1.
\end{aligned}
\end{equation}
Exact details on these are discussed in the main text.

We now discuss the fitting procedure given in \cite{a_semiparametric_spatiotemporal_Hawkes_type_point_process_model_with_periodic_background_for_crime_data} with adaption made to one-dimension and unidirectional spatial triggering.
Throughout this section, we will denote a kernel function with bandwidth $\omega$ as $z_\omega(x)$, and let $m_d, m_w$ be the number of data-points (in our case minutes) in a day and week respectively.
Then, as discussed in \cite{a_semiparametric_spatiotemporal_Hawkes_type_point_process_model_with_periodic_background_for_crime_data} and \cite{second_order_residual_analysis_of_spatiotemporal_point_processes_and_applications_in_model_evaluation}, for a spatio-temporal point process $N_p$ with conditional intensity function $\lambda(t,x)$ and a predictable process $f(t,x) \geq 0$, over a time domain $[T_1, T_2]$ and space domain $S$, we can write:
\begin{equation}\label{equ:ExpetationTheorem}
\mathbb{E}\left[ \int_{[T_1, T_2] \times S} f(t,x)dN_p(dt \times dx) \right] = \mathbb{E} \left[ \int_{T_1}^{T_2} \int_{S} f(t,x)\lambda(t,x) dxdt \right]. 
\end{equation}
Using this, we then outline the methodology provided in \cite{a_semiparametric_spatiotemporal_Hawkes_type_point_process_model_with_periodic_background_for_crime_data} to construct first the background components of our desired process then the triggering components.

\subsection{Background Components}

To construct each component our our model, we have a dataset of events, each with a time of occurrence $t_i$ and location $x_i$, with $i \in \{1, 2, ..., N\}$, and our data collected through a time-space range $[0, T] \times [0, X]$.

\subsubsection{Spatial Background}

The background spatial component, $\mu_s(x)$ can be constructed as follows. 
Denote:
\begin{equation}
\psi(t, x) = \frac{\mu_0\mu_d(t)\mu_w(t)\mu_t(t)\mu_s(x)}{\lambda(t,x)},
\end{equation}
and then substitute this into Eq.~(\ref{equ:ExpetationTheorem}), giving:
\begin{equation}\label{equ:ApproxSpatialBackground}
\begin{split}
&\sum_{i=1}^N \psi(t_i, x_i) \mathbbm{1}_{x_i \in \left[ x - \Delta x, x + \Delta x \right] } \\
&\approx \int_{0}^T \int_{0}^{X} \psi(t, \chi)\lambda(t,\chi) \mathbbm{1}_{\chi \in \left[ x - \Delta x, x + \Delta x \right] } d\chi dt \\
&= \int_{0}^T \int_{0}^{X} \frac{\mu_0\mu_d(t)\mu_w(t)\mu_t(t)\mu_s(\chi)}{\lambda(t,\chi)} \lambda(t,\chi) \mathbbm{1}_{\chi \in \left[ x - \Delta x, x + \Delta x \right] } d\chi dt \\
&= \mu_0\left( \int_{0}^T \mu_d(t)\mu_w(t)\mu_t(t) dt \right) \left( \int_{x-\Delta x}^{x+\Delta x} \mu_s(\chi) d\chi \right) \\
&\approx  \mu_0\left( \int_{0}^T \mu_d(t)\mu_w(t)\mu_t(t) dt \right) 2\mu_s(x)\Delta x \\
&\propto \mu_s(x)
\end{split}
\end{equation}

where $\Delta x$ is a small positive value and $\mathbbm{1}_{x}$ is the indicator function, giving one if its input is true and zero otherwise.
This suggests that we can write:
\begin{equation}\label{equ:HistogramEstimateSpaital}
\hat{\mu}_s(x) \propto \sum_{i=1}^N \psi_i \mathbbm{1}_{x_i \in \left[ x - \Delta x, x + \Delta x \right] }
\end{equation}
\begin{equation}\label{equ:PhiIEstiamte}
\psi_i = \frac{\mu_0\mu_d(t_i)\mu_w(t_i)\mu_t(t_i)\mu_s(x_i)}{\lambda(t_i,x_i)}.
\end{equation}
In Eq.~(\ref{equ:HistogramEstimateSpaital}), we essentially have a histogram estimate of $\mu_s(x)$, and it is natural to considering smoothing such an estimate as in the original work.
A common way to do this is through kernel smoothing, essentially smoothing a discrete data-point in time or space over a wider range of values through some kernel function, $k_\omega(x)$.
Here, we denote $\omega$ as the bandwidth of the kernel, and as commonly done specify $k$ to be Gaussian:
\begin{equation}
k_\omega(x) = \frac{1}{\sqrt{2\pi}\omega}e^{-\frac{x^2}{2\omega^2}}.
\end{equation}
Applying this smoothing, we then attain:
\begin{equation}\label{equ:OriginalEstimateSpaital}
\hat{\mu}_s(x) \propto \sum_{i=1}^N \psi_i \frac{k_{\omega_s}(x - x_i)}{\int_{0}^X k_{\omega_s}(\chi - x_i) d\chi}
\end{equation}
where $\omega_s$ is a bandwidth specific to the spatial background component, and we have prevented the `leaking of mass' problem by ensuring that each kernel function is normalized such that its integral is 1 over the domain in question.

\subsubsection{Temporal Background}

In addition to the spatial background, we also model two periodic background components, daily and weekly. 
These are constructed very similarly, so we only detail one.
To construct a periodic daily component, we take a periodic domain on $[0, m_d]$.
We then map all values inside of this domain. 
An event $t_i$ would attain a mapped value of $t_i - m_d\left\lfloor \frac{t_i}{m_d} \right\rfloor$.
The smoothed contribution of the point $t_i$ to the function $\mu_d(t)$ is then $k_{\omega_d}\left( t -  \left[ t_i - m_d\left\lfloor \frac{t_i}{m_d} \right\rfloor \right] \right)$.
However, since the domain is periodic, we account for the fact that some event $t_i$ may also have non-zero contribution from its location on the previous and following day. 
In the context of a periodic temporal function, we would consider an event before our domain as $t_i - m_d\left\lfloor \frac{t_i}{m_d} \right\rfloor - m_d$ and after our domain as $t_i - m_d\left\lfloor \frac{t_i}{m_d} \right\rfloor + m_d$.
We denote the contributions of these components as $k^{lower}_{daily} = k_{\omega_d}\left( t -  \left[ t_i - m_d\left\lfloor \frac{t_i}{m_d} \right\rfloor - m_d \right] \right)$ and $k^{upper}_{daily} = k_{\omega_d}\left( t -  \left[ t_i - m_d\left\lfloor \frac{t_i}{m_d} \right\rfloor + m_d \right] \right)$.
We can then incorporate these into the estimate as:
\begin{equation}\label{equ:OriginalEstimateDaily}
\hat{\mu}_d(t) \propto \sum_{i=1}^N w_i^{(d)} \frac{ k^{lower}_{daily} + k_{\omega_d}\left( t -  \left[ t_i - m_d\left\lfloor \frac{t_i}{m_d} \right\rfloor \right] \right) + k^{upper}_{daily}  }{ \int_{-m_d}^{2m_d} k_{\omega_d}(\tau - t_i) d\tau }
\end{equation}
with:
\begin{equation}\label{equ:w_i_d_estimate}
w_i^{(d)} = \frac{\mu_d(t_i)\mu_s(x_i)}{\lambda(t_i,x_i)}.
\end{equation}
In Eq.~(\ref{equ:OriginalEstimateDaily}) we have normalized the kernel over the entire domain, including the extended points.
Notice here that we have approximated the function over the domain $[-m_d, 2m_d]$ and then we retain only the chunk in $[0, m_d]$ for $\mu_d(t)$, meaning we do not need to provide boundary correction as there is no boundary.
We also note that one could consider an infinite sum of all previous and following days that contribute in increasingly small ways to the periodic function, however due to the choice of Gaussian kernel and the range it decays, any values further than a full period away should have negligible influence.
Finally, note that one would only input $t$ values that are in $[0, m_d]$, so functionally for any code, we would take a raw $t$ value and map it onto this domain before evaluating the formula.
Similarly, one can write the weekly periodic component as: 
\begin{equation}\label{equ:OriginalEstimateWeekly}
\hat{\mu}_w(t) \propto \sum_{i=1}^N w_i^{(w)} \frac{ k^{lower}_{weekly} + k_{\omega_w}\left( t -  \left[ t_i - m_w\left\lfloor \frac{t_i}{m_w} \right\rfloor \right] \right) + k^{upper}_{weekly}  }{ \int_{-m_w}^{2m_w} k_{\omega_w}(\tau - t_i) d\tau }
\end{equation}
with: 
\begin{equation}\label{equ:w_i_w_estimate}
w_i^{(w)} = \frac{\mu_w(t_i)\mu_s(x_i)}{\lambda(t_i,x_i)}.
\end{equation}
The same sense of a periodic function having no boundary can be accounted for in the spatial background in the same way, altering Eq.~(\ref{equ:OriginalEstimateSpaital}).

Finally, one can derive the same formulation for the background trend component as: 
\begin{equation}\label{equ:OriginalEstimateTrend}
\hat{\mu}_t(t) \propto \sum_{i=1}^N w_i^{(t)} \frac{k_{\omega_t}(t - t_i)}{\int_{0}^T k_{\omega_t}(\tau - t_i) d\tau}
\end{equation}
\begin{equation}\label{equ:w_i_t_estimate}
w_i^{(t)} = \frac{\mu_t(t_i)\mu_s(x_i)}{\lambda(t_i,x_i)}.
\end{equation}

\subsection{Triggering Components}

To determine our triggering functions in an entirely data-driven way, we first consider two data points $\left(\tau^{(1)}, \chi^{(1)}\right)$ and $\left(\tau^{(2)}, \chi^{(2)}\right)$ where $\tau^{(1)} < \tau^{(2)}$ and $ \chi^{(1)} > \chi^{(2)}$.
We then define $\rho(\tau^{(1)}, \chi^{(1)}, \tau^{(2)}, \chi^{(2)})$ as:
\begin{equation}
\rho\left(\tau^{(1)}, \chi^{(1)}, \tau^{(2)}, \chi^{(2)}\right) = \begin{cases}
      \frac{A g( \tau^{(2)} - \tau^{(1)} )h( \chi^{(2)} - \chi^{(1)} )}{\lambda(\tau^{(2)}, \chi^{(2)})}, & \text{if} \, \, \tau^{(1)} < \tau^{(2)} \, \, \text{and} \, \, \chi^{(1)} > \chi^{(2)} \\
      0, & \text{otherwise}
    \end{cases}
\end{equation}

If we apply this in Eq.~(\ref{equ:ExpetationTheorem}), letting $f(\tau, \chi) = \rho(t_i, x_i, \tau, \chi)\mathbbm{1}_{ \tau - t_i \in [t - \Delta t, t + \Delta t] }$, then we attain:
\begin{equation}\label{equ:DeriveTemporalTrigger}
\begin{split}
&\sum_{j}\rho(t_i, x_i, \tau_j, \chi_j)\mathbbm{1}_{ \tau_j - t_i \in [t - \Delta t, t + \Delta t] } \\
&\approx \int_{0}^T \int_{0}^X \rho(t_i, x_i, \tau, \chi)\mathbbm{1}_{ \tau - t_i \in [t - \Delta t, t + \Delta t] } \lambda(\tau, \chi) d\chi d\tau \\
&= A\int_{t_i}^T \int_{0}^{x_i}  \frac{g(\tau - t_i) h(\chi - x_i)}{\lambda(\tau, \chi)} \mathbbm{1}_{ \tau - t_i \in [t - \Delta t, t + \Delta t] } \lambda(\tau, \chi) d\chi d\tau \\
&= A\left(\int_{t_i}^T g(\tau - t_i)\mathbbm{1}_{ \tau - t_i \in [t - \Delta t, t + \Delta t] }d\tau\right) \left( \int_{0}^{x_i} h(\chi - x_i) d\chi \right).
\end{split}
\end{equation} 
If we then let $s = \tau - t_i$ we have:
\begin{equation}
\begin{split}
&\sum_{j}\rho(t_i, x_i, \tau_j, \chi_j)\mathbbm{1}_{ \tau_j - t_i \in [t - \Delta t, t + \Delta t] } \\
&\approx A\left(\int_{0}^{T-t_i} g(s)\mathbbm{1}_{ s \in [t - \Delta t, t + \Delta t] }ds\right) \left( \int_{0}^{x_i} h(\chi - x_i) d\chi \right) \\
&= A\left(\int_{t - \Delta t}^{t + \Delta t} g(s)ds\right) \left( \int_{0}^{x_i} h(\chi - x_i) d\chi \right) \\ 
&\approx 2A\Delta t g(t) \left( \int_{0}^{x_i} h(\chi - x_i) d\chi \right)\\
&\propto g(t). 
\end{split}
\end{equation}

As a result:
\begin{equation}\label{equ:TemporalTriggerHistogramPart1}
\sum_i \sum_j \rho(t_i, x_i, t_j, x_j)\mathbbm{1}_{ t_j - t_i \in [t - \Delta t, t + \Delta t] } \propto g(t)
\end{equation}
and hence:
\begin{equation}\label{equ:TemporalTriggerHistogram}
\hat{g}(t) \propto \sum_{(i, j)}  \rho_{i,j}\mathbbm{1}_{ t_j - t_i \in [t - \Delta t, t + \Delta t] }
\end{equation}
with:
\begin{equation}\label{equ:rhoij}
\rho_{i,j} = \frac{Ag(t_j-t_i)h(x_j-x_i)}{\lambda(t_j,x_j)}, \, \,  \text{for all} \, \, (i,j) \ \text{s.t.}\ t_i < t_j \ \text{and}\ x_i > x_j. 
\end{equation}
As before, the estimator in Eq.~(\ref{equ:TemporalTriggerHistogram}) can be smoothed to give:
\begin{equation}\label{equ:OriginalTemporalTrigger}
\hat{g}(t) \propto \frac{ \sum\limits_{(i, j)}  \frac{\rho_{i,j}k_{\omega_g}\left( t - t_j + t_i \right)}{\int_0^{T-t_i} k_{\omega_g}\left( \tau - t_j + t_i \right) d\tau } }{ \sum\limits_{i=1}^N \mathbbm{1}_{t_i + t \leq T} }.
\end{equation}
In Eq.~(\ref{equ:OriginalTemporalTrigger}), we have normalized each kernel density estimate by the remaining time, and then finally dividing by the term $\sum_{i=1}^N \mathbbm{1}_{t_i + t \leq T}$, which corrects for repetitions as detailed in \cite{a_semiparametric_spatiotemporal_Hawkes_type_point_process_model_with_periodic_background_for_crime_data}.

In the same way, we can write the spatial triggering function as:
\begin{equation}\label{equ:OriginalSpatialTrigger}
\hat{h}(x) \propto \frac{ \sum\limits_{(i, j)}  \frac{\rho_{i,j}k_{\omega_h}\left( x - x_j + x_i \right)}{\int_{0}^{x_i} k_{\omega_h}\left( \chi - x_j + x_i \right) d\chi } }{ \sum\limits_{i=1}^N \mathbbm{1}_{x_i + x \geq 0} }.
\end{equation}
where the scaling integral has been adapted to account for our one-directional triggering in space.

\subsection{Determining Coefficients $A, \mu_0$}\label{sec:DetermineAMu0}

In the model specification, we have two coefficients $A$ and $\mu_0$ that specify the how much or little of the triggering and background component respectively enters the conditional intensity function.
In-particular, it should be noted that $A$ can be interpreted as the proportion of the impact of the triggering function on the total intensity, so a high $A$ value suggests data is dominated by triggering and the converse for a small value.
These parameters can be determined through maximum likelihood estimation.
The log-likelihood function for a spatial-temporal point process model with triggering is given by:
\begin{equation}\label{equ:LogLikelihood}
\log(L) = \sum_{i=1}^{N} \log\left( \lambda(t_i, x_i) \right) - \int_{0}^T\int_{0}^X \lambda(t,x) dxdt
\end{equation}
Using Eq.~(\ref{equ:BasicModelEquationSupp}), we then see:
\begin{equation}
\begin{split}
\log(L) &= \begin{multlined}[t] \sum_{i=1}^N \log \left( \mu_0\mu_d(t)\mu_w(t)\mu_t(t)\mu_s(x) + A\sum_{\substack{t_i < t \\ x_i > x}}g(t-t_i)h(x-x_i) \right) - \\ \int_{0}^T\int_{0}^X \mu_0\mu_d(t)\mu_w(t)\mu_t(t)\mu_s(x) + A\sum_{\substack{t_i < t \\ x_i > x}}g(t-t_i)h(x-x_i) dx dt 
\end{multlined}\\
&= \begin{multlined}[t] \sum_{i=1}^N \log \left( \mu_0\mu_d(t)\mu_w(t)\mu_t(t)\mu_s(x) + A\sum_{\substack{t_i < t \\ x_i > x}}g(t-t_i)h(x-x_i) \right) - \\ \int_{0}^T\int_{0}^X \mu_0\mu_d(t)\mu_w(t)\mu_t(t)\mu_s(x) dx dt - \int_{0}^T\int_{0}^X A\sum_{\substack{t_i < t \\ x_i > x}}g(t-t_i)h(x-x_i) dx dt 
\end{multlined}\\
&= \begin{multlined}[t] \sum_{i=1}^N \log \left( \mu_0\mu_d(t)\mu_w(t)\mu_t(t)\mu_s(x) + A\sum_{\substack{t_i < t \\ x_i > x}}g(t-t_i)h(x-x_i) \right) - \\ \mu_0\int_{0}^T\int_{0}^X \mu_d(t)\mu_w(t)\mu_t(t)\mu_s(x) dx dt - A \sum_{i=1}^N\int_{t_i}^T\int_{0}^{x_i} g(t-t_i)h(x-x_i) dx dt. 
\end{multlined}
\end{split}
\end{equation}
We then denote:
\begin{equation}
\begin{split}
U &= \int_{0}^T\int_{0}^X \mu_d(t)\mu_w(t)\mu_t(t)\mu_s(x) dx dt \\
G &= \sum_{i=1}^N\int_{t_i}^T\int_{0}^{x_i} g(t-t_i)h(x-x_i) dx dt
\end{split}
\end{equation}
and attain the partial derivatives with respect to $A$ and $\mu_0$ as:
\begin{equation}
\begin{split}
\frac{\partial \log(L)}{\partial \mu_0}     &= \sum_{i=1}^N \frac{\mu_d(t_i)\mu_w(t_i)\mu_t(t)\mu_s(x)}{\lambda(t_i, x_i)} - U \\
\frac{\partial \log(L)}{\partial A} &= \sum_{i=1}^N \frac{ \sum\limits_{\substack{t_i < t \\ x_i > x}}g(t_j-t_i)h(x_j-x_i) }{\lambda(t_i, x_i)} - G.
\end{split}
\end{equation}
Setting these equal to 0 and solving, one attains the following iterative system of equations:
\begin{equation}\label{equ:IterSystem}
\begin{split}
\psi_i^{(\zeta)} &= \frac{\mu_0^{(\zeta)}\mu_d(t_i)\mu_w(t_i)\mu_t(t_i)\mu_s(x_i)}{\mu_0^{(\zeta)}\mu_d(t_i)\mu_w(t_i)\mu_t(t_i)\mu_s(x_i) + A^{(\zeta)} \sum\limits_{\substack{t_j < t_i \\ x_j > x_i}}g(t_i-t_j)h(x_i-x_j) } \\
A^{\zeta+1} &= \frac{N - \sum\limits_{i=1}^N \psi_i^{(\zeta)}}{G} \\
\mu_0^{\zeta+1} &= \frac{N - A^{(\zeta+1)}G}{U}. \\
\end{split}
\end{equation}

\subsection{Fitting Algorithm}\label{sec:Fitting}

Our final reference back to the original work \cite{a_semiparametric_spatiotemporal_Hawkes_type_point_process_model_with_periodic_background_for_crime_data} is the description of the fitting algorithm used, with our additional components incorporated, detailed in Algorithm \ref{alg:ModelFitting}.
\RestyleAlgo{boxruled}
\LinesNumbered
\begin{algorithm}[ht]
	\caption{Fit Triggering Point Process Model to Data\label{alg:ModelFitting}}
	\KwIn{Initial guesses $\mu_d, \mu_w, \mu_t, \mu_s, g, h, A, \mu_0$}
	\While{Not converged}{
	Compute $w_i^{(d)}, w_i^{(w)}, w_i^{(t)}, \psi_i, \rho_{i,j}$ for all valid $(i,j)$ using Eq. (\ref{equ:w_i_d_estimate}), (\ref{equ:w_i_w_estimate}), (\ref{equ:w_i_t_estimate}), (\ref{equ:PhiIEstiamte}) and (\ref{equ:rhoij}) \\
	Estimate $\mu_d, \mu_w, \mu_t, \mu_s, g, h$ using Eq. (\ref{equ:OriginalEstimateDaily}), (\ref{equ:OriginalEstimateWeekly}), (\ref{equ:OriginalEstimateTrend}), (\ref{equ:OriginalEstimateSpaital}), (\ref{equ:OriginalTemporalTrigger}) and (\ref{equ:OriginalSpatialTrigger}). Appropriate boundary correction should be applied if desired. \\
	Determine $p_i \, \, \, \text{for all} \, \, i \in [1, 2, \dots N]$ by solving Eq. (\ref{equ:OptimProblemMonotonicSupp}) \\
	Estimate $A$ and $\mu_0$ using Eq. (\ref{equ:IterSystem}) 
	}
	\KwOut{ Optimized components $\mu_d, \mu_w, \mu_t, \mu_s, g, h, A, \mu_0$}    
\end{algorithm}

For computational speed, we assume triggering after 12 hours is 0, which is informed by considering time-scales similar to the worst recorded traffic jams in the U.K, detailed in \cite{inrix_worst_traffic_jams}.
For those interested in an implimentation of the original methodology detailed in \cite{a_semiparametric_spatiotemporal_Hawkes_type_point_process_model_with_periodic_background_for_crime_data}, this can be found at \url{https://rss.onlinelibrary.wiley.com/hub/journal/1467985x/series-a-datasets}.

\section{Examples of Boundary Correction}

\begin{figure}[!ht]
    \centering
    \subfloat[][Mirrored correction example, using data generated from a half normal distribution with mean 0, standard deviation 2. Each density estimate is fit with a bandwidth of 0.75.]{\includegraphics[width=0.48\textwidth]{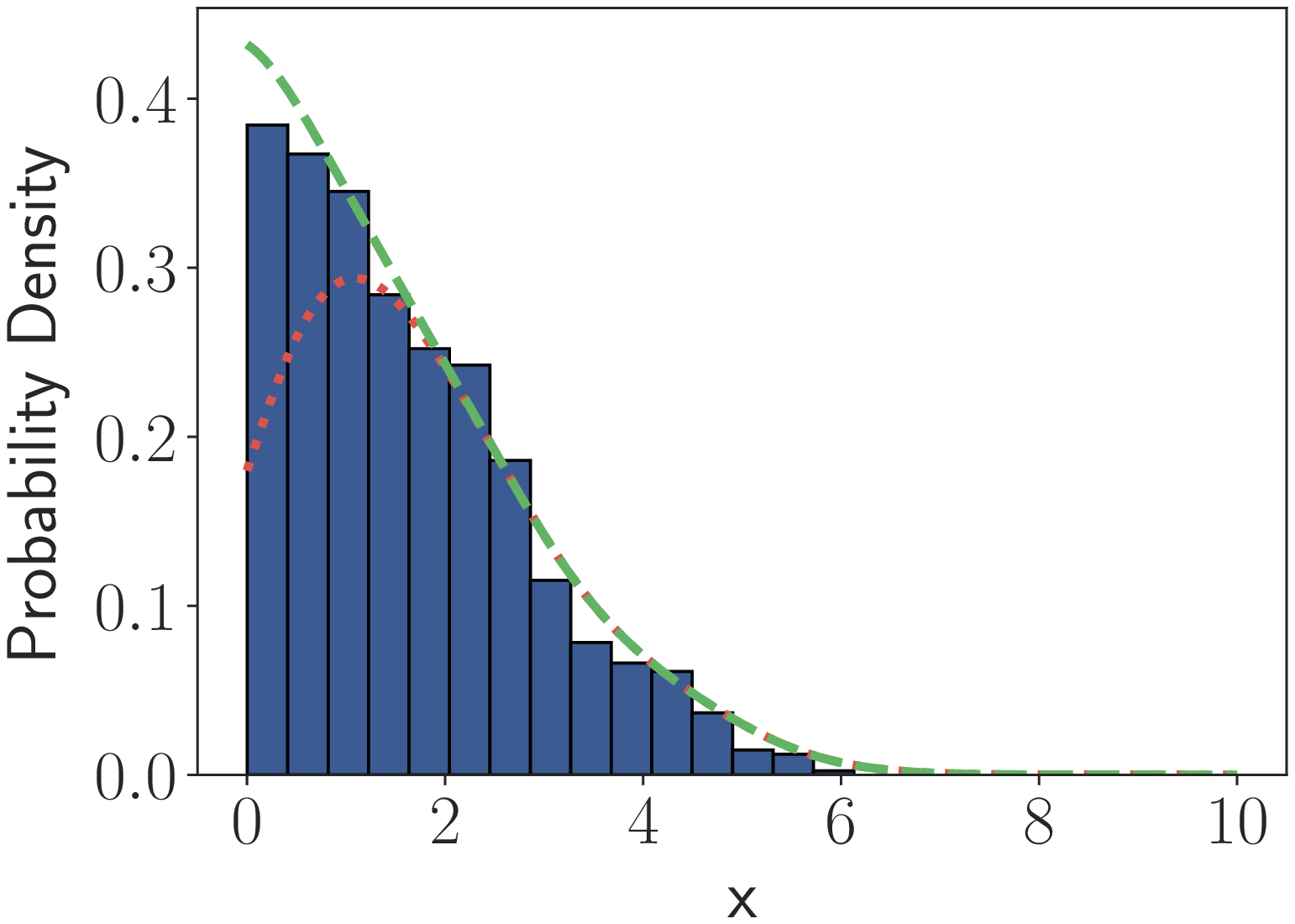}\label{fig:MirroredExample}}
    \hfill
    \subfloat[][Periodic correction example, using data generated from a gamma distribution with shape 5 and scale 1. Each density estimate is fit with a bandwidth of 0.5.]{\includegraphics[width=0.48\textwidth]{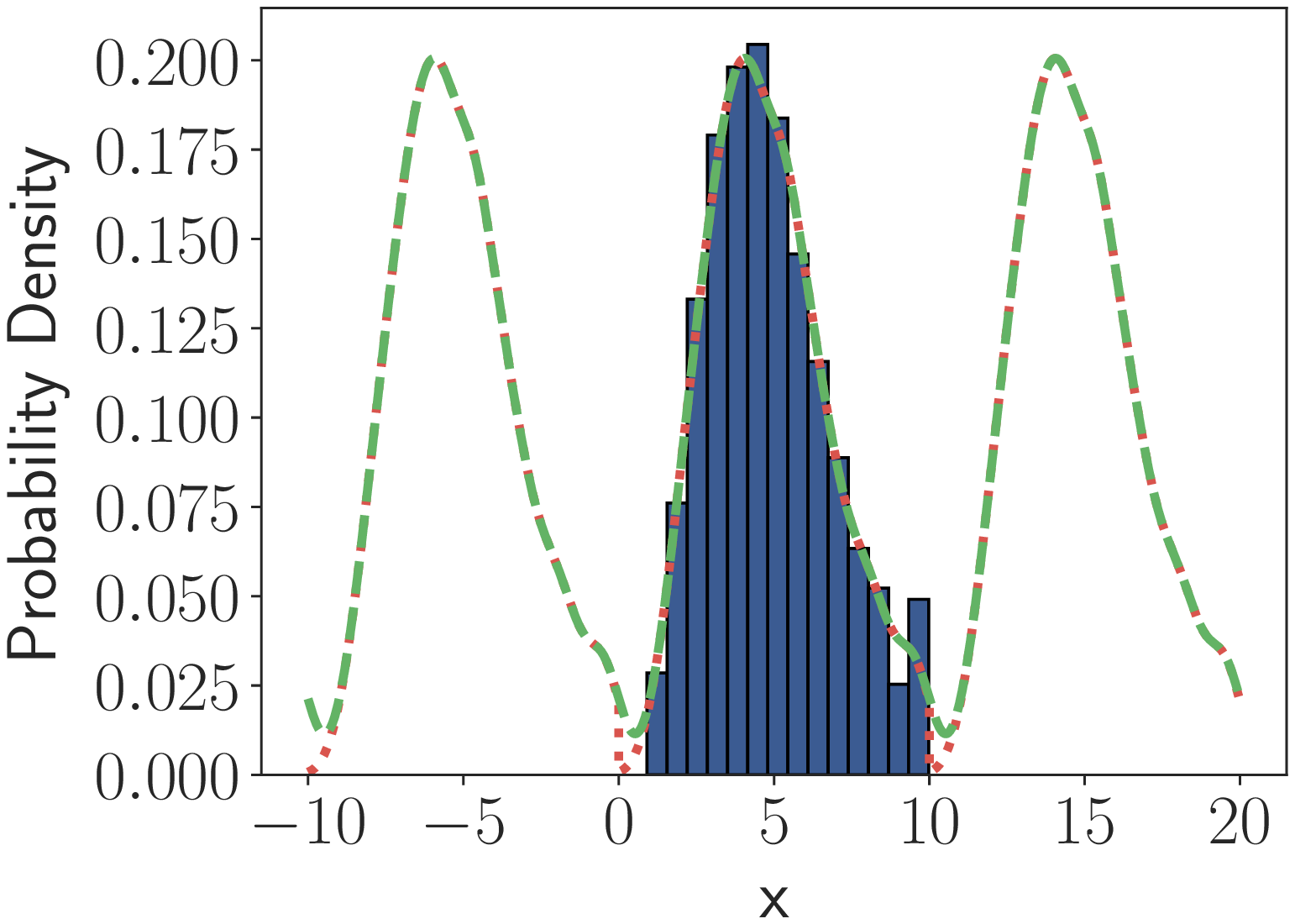}\label{fig:PeriodicExample}}

    \caption{Examples of boundary correction influence on simulated data. We show the data as a histogram, the non-corrected fits with ({{\PaleRedDotted}}) and the boundary corrected fits with ({{\BoringGreenDashed}}).}\label{fig:SimulatedCorrectionExamples}
\end{figure}

In the main text and in the previous section to this, we discussed methods of boundary correction.
Specifically, we use mirrored correction when the domain is truncated, and for periodic domains we include influence from points lying one period either side of the domain in question. 
It can be useful to visualize what impact such methods are actually having on an estimate, so we generate simulated data to show exactly how they impact the estimates.
In Fig. \ref{fig:SimulatedCorrectionExamples}, we show examples of two datasets. 
In Fig. \ref{fig:MirroredExample}, we show a data-set truncated at 0, and fit two density estimates to it, one with and one without mirrored boundary correction. 
In Fig. \ref{fig:PeriodicExample}, we show a dataset generated on the domain $[0,10]$, and imagine that we are trying to construct a periodic function from it, with $[0,10]$ being one period. 
Again, we show an example of a density fit without any boundary correction, and one with the additional periodic components included.

Inspecting Fig. \ref{fig:MirroredExample}, we see a clear drop in the estimate at the truncation point of 0 without correction.
Additionally, from Fig. \ref{fig:PeriodicExample}, we see again a drop at the two ends of the periodic domain without any correction.
These are removed when we apply our correction methods.

\section{Temporal Background Analysis Around Peaks in Spatial Intensity}\label{HotspotComparision}

\begin{figure}[!ht]
    \centering
    \subfloat[][Visualization of hotspot locations.]{\includegraphics[width=0.32\textwidth]{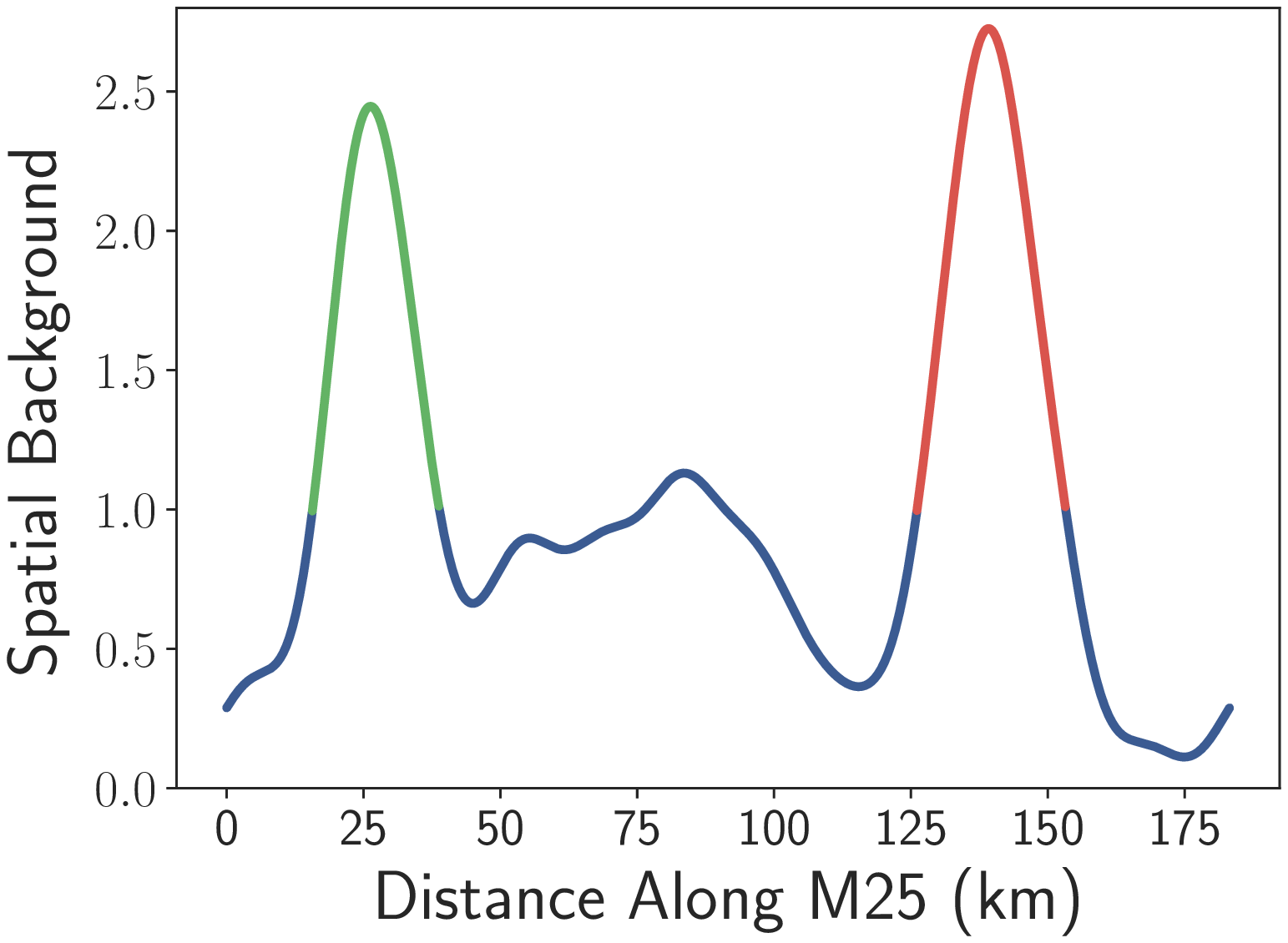}\label{fig:HotspotComarisionLocations}}
    \hfill
    \subfloat[][Comparison of daily backgrounds.]{\includegraphics[width=0.32\textwidth]{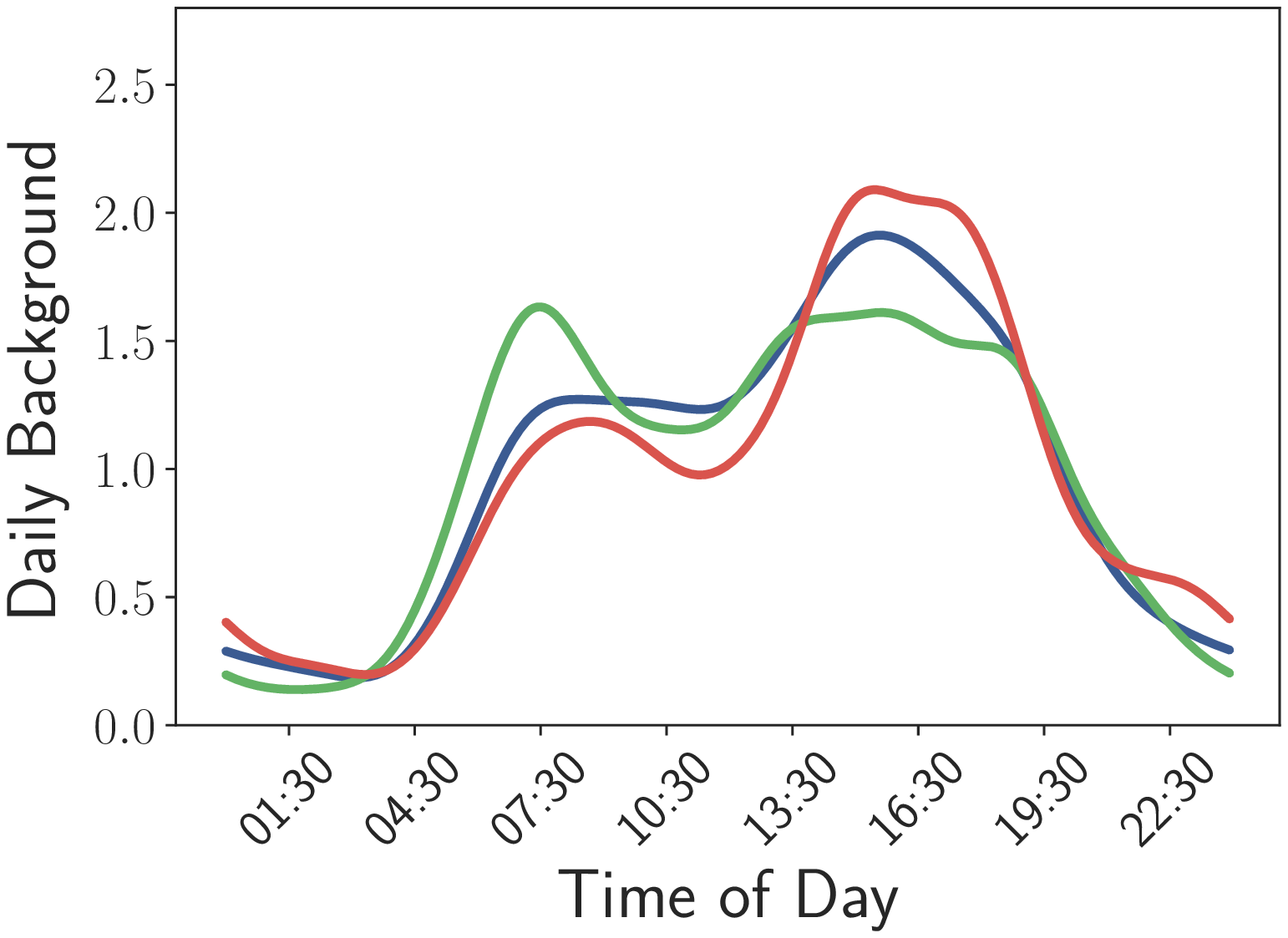}\label{fig:DailyBackgroundHotspotCompare}}
    \hfill
    \subfloat[][Comparison of weekly backgrounds.]{\includegraphics[width=0.32\textwidth]{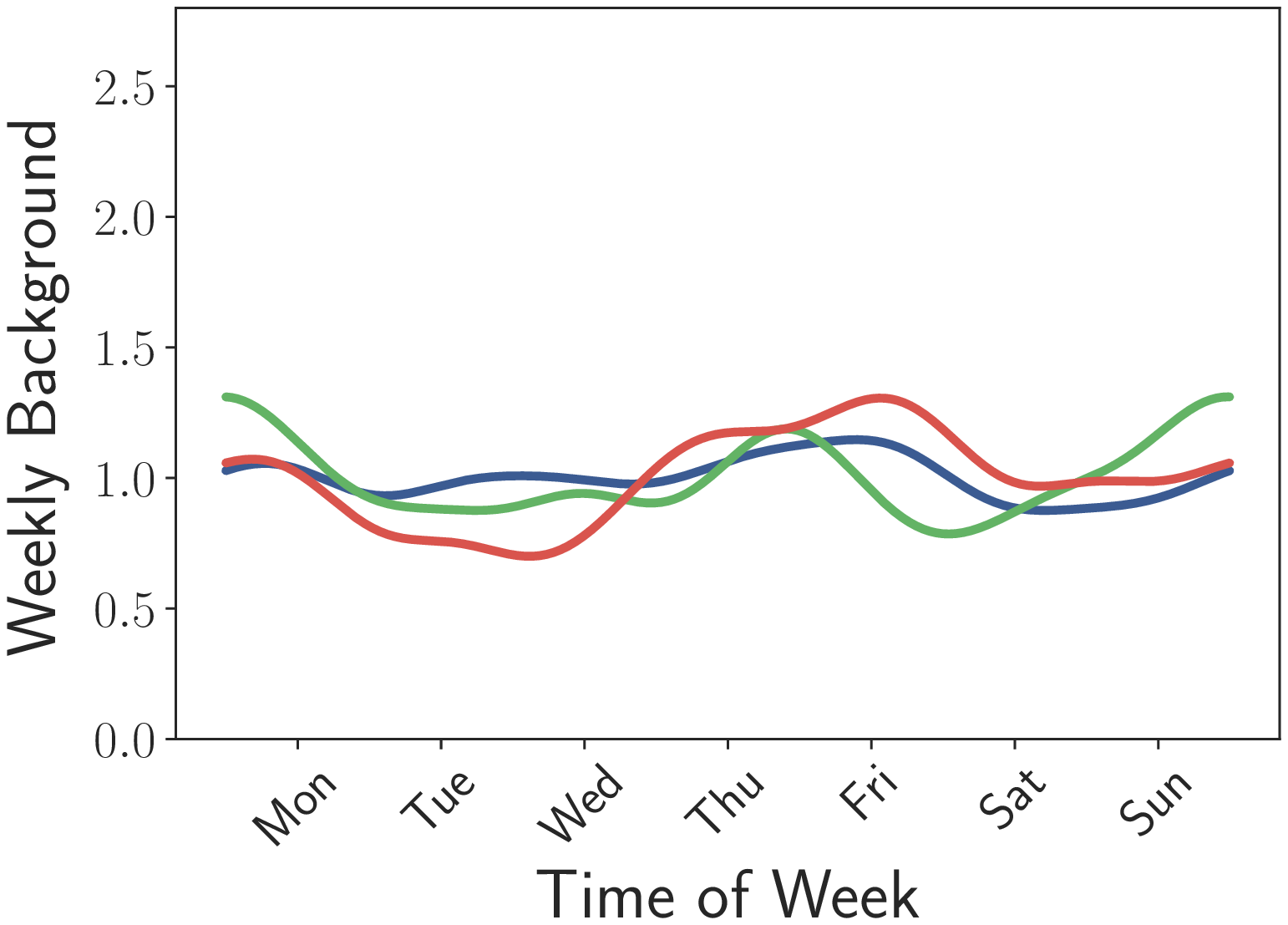}\label{fig:WeeklyBackgroundHotspotCompare}}
    \caption{Identification of hotspot locations, and comparison of temporal background components around them. In \ref{fig:HotspotComarisionLocations}, we show the spatial locations of hotspot 1 ({{\BoringGreenFull}}) and hotspot 2 ({{\PaleRedFull}}). In \ref{fig:DailyBackgroundHotspotCompare} and \ref{fig:WeeklyBackgroundHotspotCompare}, we show the results for the entire dataset ({{\DenimBlueFull}}), hotspot 1 ({{\BoringGreenFull}}) and hotspot 2 ({{\PaleRedFull}}). }\label{fig:HotspotComarision}
\end{figure}

Whilst we have seen that the background component of our model has two clear peaks in it, we can also question if the daily and weekly background components differ significantly in the vicinity of these peaks, compared to their behaviour across the entire motorway.
To investigate this, we consider each peak separately.
Firstly, we define two `hotspots' surrounding the two peaks in spatial intensity.
These locations start where the spatial background component has a value above 1, and end where it then falls back below 1. 
We then isolate a set of events that occour in each spatial hotspot. 
For the form of our model, this suggests we are isolating the spatial location where there is a increase in the rate of events compared to the average spatial location.
We then re-fit our model using only this subset of events, and question what resulting daily and weekly background components arise.
We visualize our results in Fig. \ref{fig:HotspotComarision}.
Specifically, we visualize the subsets by location in Fig. \ref{fig:HotspotComarisionLocations}, the daily background in Fig. \ref{fig:DailyBackgroundHotspotCompare} and the weekly background in \ref{fig:WeeklyBackgroundHotspotCompare}.

From Fig. \ref{fig:HotspotComarision}, it is clear that the temporal background around each hotspot is reasonably similar to that across the entire M25.
The daily background around the first hotspot has a slightly more pronounced peak in the morning and lower peak in the evening compared to the entire network.
The second hotspot has a slightly higher peak in the evening.
Both hotspots have a similar weekly component.
Overall, it seems reasonable to conclude that behaviour at these two hotspots is not fundamentally different to that across the entire motorway.

\bibliographystyle{rss}
\setlength{\bibsep}{0.05pt}
\bibliography{references}

\end{document}